\documentclass[journal=jacsat,manuscript=article]{achemso}
\usepackage{achemso}
\usepackage[version=3]{mhchem}
\usepackage[english]{babel}
\usepackage{filecontents}
\usepackage{graphicx}
\usepackage{courier}
\usepackage{amssymb,amsmath,amsthm,mathrsfs,comment,afterpage,accents}
\usepackage{mathtools}
\usepackage{braket}
\usepackage{mathrsfs}
\usepackage{courier}
\usepackage{color}
\usepackage{subdepth}
\usepackage[graphicx]{realboxes}
\usepackage{rotating}
\setkeys{acs}{articletitle = true}

\makeatletter
\def\@dotsep{4.5}
\makeatother
\usepackage{dcolumn}
\usepackage{bm}
\renewcommand\vec\mathbf
\newcommand\mat\mathbf

\setkeys{acs}{maxauthors = 0}
\newcommand{\insertnew}[1]{{\textcolor{black} {#1}}}
\newcommand{\revision}[1]{{\textcolor{black} {#1}}}
\newcommand{\replacewith}[2]{{\textcolor{red}{}}{\textcolor{black}{#2}}}

\title{Regularized Orbital-Optimized Second-Order M{\o}ller-Plesset Perturbation Theory: A Reliable Fifth-Order Scaling Electron Correlation Model with Orbital Energy Dependent Regularizers
}

\author{Joonho Lee}
\email{linusjoonho@gmail.com}
\author{Martin Head-Gordon}
\email{mhg@cchem.berkeley.edu}
\affiliation{
Department of Chemistry, University of California, Berkeley, California 94720, USA
Chemical Sciences Division, Lawrence Berkeley National Laboratory, Berkeley, California 94720, USA
}
\begin{document}
\newpage
\maketitle
\begin{abstract}
We derive and assess two new classes of regularizers that cope with offending denominators in the single-reference second-order M{\o}ller-Plesset perturbation theory (MP2). In particular, we discuss the use of two types of orbital energy dependent regularizers, $\kappa$ and $\sigma$, in conjunction with orbital-optimized MP2 (OOMP2). The resulting fifth-order scaling methods, $\kappa$-OOMP2 and $\sigma$-OOMP2, have been examined for bond-breaking, thermochemistry, non-bonded interactions and biradical problems. Both methods with strong enough regularization restore restricted to unrestricted instability (i.e. Coulson-Fischer points) \insertnew{that unregularized OOMP2 lacks} when breaking \insertnew{bonds in} \ce{H2}, \ce{C2H6}, \ce{C2H4}, \insertnew{and} \ce{C2H2}. The training of \insertnew{the $\kappa$ and $\sigma$} regularization parameters was performed with the W4-11 set. We further developed scaled correlation energy variants, $\kappa$-S-OOMP2 and $\sigma$-S-OOMP2, by training on the TAE140 subset of the W4-11 set. Those new OOMP2 methods were tested on the RSE43 set and the \replacewith{TA14}{TA13} set where unmodified OOMP2 itself performs very well. The modifications we made were found insignificant in these data sets. Furthermore, we tested the new OOMP2 methods on singlet biradicaloids using Yamaguchi's approximate spin-projection. Unlike the unregularized OOMP2, which fails to converge these systems due to the singularity, we show that regularized OOMP2 methods successfully capture strong biradicaloid characters. 
\insertnew{While further assessment on larger datasets is desirable, $\kappa$-OOMP2 with $\kappa$ = 1.45 $E_{h}^{-1}$ appears to combine favorable recovery of Coulson-Fischer points with good numerical performance.}
\end{abstract}
\newpage
\section{Introduction}
The single-reference second-order M{\o}ller-Plesset perturbtation theory (MP2) is one of the simplest correlated wavefunction methods (and therefore one of the most popular ones). There have been some significant developments in improving different aspects of MP2 in the past decade or so and we shall mention those that are particularly relevant to this work.

The development of the resolution-of-identity (RI) technique (or the density-fitting technique) for MP2 was revolutionary.\cite{Feyereisen1993,Bernholdt1996} Although RI-MP2 has fundamentally the same computational scaling as MP2 (i.e. $\mathcal{O}(N^5)$), it substantially reduces the prefactor of the algorithm and has allowed for large-scale applications of MP2. RI-MP2 is now considered the {\it de facto} algorithm for any MP2 calculations except for systems with off-atom electrons such as \insertnew{dipole-bound electrons\cite{Jordan2003} or electronic resonances.\cite{Morgan1981}}
Given its popularity, we shall focus on building a new theory on top of RI-MP2 and we will refer RI-MP2 to as just MP2 for simplicity for the following discussion. 

Aside from faster MP2 algorithms, there are two common ways to improve the energetics of MP2: one is the spin component scaled (SCS)-MP2 approach \cite{Grimme2003,Jung2004,Grimme*2005,Lochan2005,DistasioJR.2007,Lochan2007a} and another is the orbital-optimized MP2 (OOMP2) method.\cite{Lochan2007,Neese2009,Bozkaya2011} 
SCS-MP2 improved the energetics of MP2 for thermochemistry and non-covalent interactions \insertnew{although the optimal scaling parameters are different for these two classes of relative energies}. 
From an efficiency standpoint, the scaled opposite-spin MP2 (SOS-MP2) method in this category is noteworthy as it is an overall $\mathcal{O}(N^4)$ algorithm while improving the energetics.\cite{Jung2004, Lochan2005,Lochan2007a} 
The idea of SCS-MP2 is also often used in double-hybrid density functional approximations. \cite{Grimme2006,Chai2009,Kozuch2011,Mardirossian2018}
\insertnew{Additionally, overbinding molecular interactions due to inherent errors in MP2 and basis set superposition error were reduced by an attenuated MP2 approach.\cite{Goldey2012,Goldey2013,Goldey2014,Goldey2015}}

OOMP2 often produces a qualitatively better set of orbitals for systems where unrestricted Hartree-Fock (UHF) orbitals exhibit artificial spin symmetry breaking. Artificial symmetry breaking is a quite common problem in open-shell systems and polyaromatic hydrocarbons that are weakly correlated systems.\cite{Davidson1983,Paldus2007} 
In such cases, using UHF orbitals for correlated wave function calculations leads to catastrophically wrong energies and properties.\cite{Farnell1983,Nobes1987,Gill1988,Jensen1990}
The use of Br{\"u}ckner orbitals often improves the results significantly, though obtaining those orbitals is quite expensive. \cite{Dykstra1977, Handy1989}
Therefore, OOMP2 was proposed as an economical way to approximate Br{\"u}ckner orbitals. \cite{Lochan2007}
Orbital optimizing at the MP2 level often restores the spin symmetry and results in far better energetics. \cite{Lochan2007,Neese2009,Soydas2015} Furthermore, OOMP2, in principle, removes the discontinuity in the first-order properties that can be catastrophic at the onset of symmetry breaking in MP2.\cite{Kurlancheek2009} 
These two observations motivated several research groups to apply \cite{Stuck2011,Kurlancheek2012} and to develop OOMP2 and its variants.\cite{Soydas2013,Bozkaya2013b,Stuck2013,Peverati2013,Sancho-Garcia2016,Bozkaya2016,Razban2017,Najibi2018}
\revision{It was also extended to higher order perturbation theory methods, such as OOMP3 and OOMP2.5.\cite{Bozkaya2011a, Bozkaya2014, Bozkaya2016a}}
\revision{The analytic nuclear gradient of OOMP2 was also efficiently implemented \cite{Bozkaya2013a, Kurlancheek2012, Bozkaya2014a}
and the Cholesky decomposition was also used for an efficient implementation. \cite{Bozkaya2014b}}

However, OOMP2 has shown multiple problems that limit its applicability. First, the inclusion of the MP2 form of the correlation energy in orbital optimization tends to produce very small energy denominators. 
 In some cases, this leads to a divergence of the total energy and it is commonly observed when stretching bonds. Moreover, this is the cause for the significant underestimation of harmonic frequencies at equilibrium geometries.\cite{Stuck2013} 
Given that it is very unlikely to observe vanishing energy denominators in finite systems at the HF level, \cite{Bach1994} the applicability of MP2 on top of HF orbitals is greater than that of OOMP2. 

Second, OOMP2 often does not continuously break spin-symmetry even when there exists a broken-symmetry solution that is lower in energy. \cite{Sharada2015} 
To have a continuous transition from a restricted (R) solution to a unrestricted (U) solution, there should be a point at which the lowest eigenvalue of the R to U stability Hessian becomes zero. In the case where two solutions are separated by a barrier, we often observe a discontinuous transition from R to U (or even no transition at all) and there are only positive non-zero eigenvalues in the R to U stability Hessian. 
We have observed multiple systems where ROOMP2 does not undergo a continuous transition to UOOMP2 while the corresponding HF calculation does. In this case, this is an artifact of OOMP2 and it is necessary to resolve this issue to reach a proper dissociation limit.

Our group has attempted to solve these two separate issues using a simple regularization scheme that shifts the energy denominator by a constant $\delta$.\cite{Stuck2013} Despite its simple form, it was effective enough to solve those two issues described above. It is immediately obvious that the energy can no longer diverge. Moreover, as the MP2 energy is damped away, the qualitative behavior of OOMP2 approaches that of HF where we observe continuous transitions from R to U. Razban et al\cite{Razban2017} tried to find a regularization parameter $\delta$ that solves those two issues within OO-SOS-MP2. Since the desired regularization strength to restore Coulson-Fischer points \cite{Coulson1949} was too strong (i.e. the true correlation energy attenuated), one faces difficulties in dealing with rather easier problems for MP2, such as typical thermochemistry problems. 
This led us to considering alternative forms of regularizers which may depend on the orbital energy gap. 
Ideally, we need a regularizer that leaves the energy contribution from a large denominator unchanged and damps away the contribution from an offending small denominator. 

We note that the idea of regularizing perturbation theory has been explored by several others. Other ideas include the level-shifted complete active space second-order perturbation theory (CASPT2),\cite{Roos1995} restrained denominator MP2 (RD-MP2),\cite{Ohnishi2014} and the recently introduced driven similarity renormalization group (DSRG) methods.\cite{Evangelista2014} In particular the DSRG methods are particularly interesting as they regularize each term differently depending on the associated energy denominator. In fact the regularizers we study here were motivated by DSRG. We also mention that there are other approaches that are computationally as simple as (or cheaper than) MP2 and do not diverge even for metallic systems. These include direct random phase approximation\cite{Eshuis2010} and degeneracy-corrected perturbation theory. \cite{Assfeld1995} \insertnew{We also mention our group's previous work on penalty functions which regularize PT amplitudes.\cite{Lawler2008}}
\revision{It is worthwhile to mention that there may be formal connections between regularized OOMP2 and correlation theories with screened interactions such as random phase approximation \cite{Chen2017} and coupled-pair theories \cite{Bartlett1978} as these all exhibit no singular behavior for metallic systems.}

This paper is organized as follows: (1) we review OOMP2 , (2) we describe the regularizers that we used in this work and derive the pertinent orbital gradient of them, (3) we investigate the effect of regularizers on the stability Hessian, (4) we demonstrate preliminary training and test of the new regularizers on the W4-11 set\cite{Karton2011}, the RSE43 set,\cite{Goerigk2017, Zipse2006} and the TA13 set,\cite{Tentscher2013} and (5) we apply these new methods to two chemically interesting biradical molecules.\cite{Hu2015,Kobayashi2017}

\section{Theory}
We will use $i,j,k,l,\cdot\cdot\cdot$ to index occupied orbitals, $a,b,c,d,\cdot\cdot\cdot$ to index virtual orbitals, and $p,q,r,s,\cdot\cdot\cdot$ to index either of those two. 

\subsection{OOMP2 Lagrangian in Spin-Orbital Basis}
We review the OOMP2 Lagrangian formulation and its orbital gradient within the spin-orbital notation. For generality, we do not assume orbitals to be real. 
\subsubsection{Hylleraas Functional}
The Hylleraas functional $J_H$ is given as
\begin{align}\nonumber
J_H[\Psi_1] & = 
\Bra{\Psi_1}\hat{F} - E_0\Ket{\Psi_1}
+ \Bra{\Psi_0}\hat{V}\Ket{\Psi_1} + \Bra{\Psi_1}\hat{V}\Ket{\Psi_0}\\
&=\Bra{\Psi_1}\hat{F}_N\Ket{\Psi_1}
+ \Bra{\Psi_0}\hat{V}_N\Ket{\Psi_1} + \Bra{\Psi_1}\hat{V}_N\Ket{\Psi_0}
,
\end{align}
where the subscript $N$ denotes ``normal-ordered'' operators \cite{Shavitt2009} and the OOMP2 ansatz for $\Ket{\Psi_1}$ by definition includes only doubly excited determinants:
\begin{equation}
\Ket{\Psi_1} = \hat{T}_2 \Ket{\Psi_0} = \frac14 \sum_{ijab} t^{ab}_{ij}\Ket{\Psi_{ij}^{ab}}.
\end{equation}
In a simpler notation, this functional is
\begin{equation}
J_H[\mathbf t] = \mathbf{t}^\dagger \Delta \mathbf{t} + \mathbf{t}^\dagger \mathbf{V} + \mathbf{V}^\dagger \mathbf{t}
\label{eq:hylleraas}
\end{equation}
where $\Delta$ is a rank-8 tensor defined as
\begin{equation}
\Delta_{ijab}^{klcd} = \langle\Psi_{ij}^{ab}|\hat{F}_N|\Psi_{kl}^{cd}\rangle
\end{equation}
and $\mathbf V$ is 
\begin{equation}
V_{ab}^{ij} = \langle ab || ij\rangle
\end{equation}
In passing we note that when we include orbital optimization effects $\Ket{\Psi_1}$ is no longer composed of canonical orbitals. 
\insertnew{Instead it is convenient to use pseudocanonical orbitals\cite{Hillier1970,Hillier1970a} that diagonalize the occupied-occupied (OO) and the virtual-virtual (VV) blocks of the Fock operator, $\hat{F}$.} 
Strictly speaking, singles contributions do not vanish. However, we argue that this is a part of our ans{\"a}tz, consistent with the idea of variational Br{\"u}ckner orbitals. Orbital optimization incorporates the most important singles effects. Indeed, the effect of singles was examined in the context of OOMP2 and found negligible as in ref. \citenum{Neese2009}.
\subsubsection{MP2 Lagrangian}
We construct a Lagrangian that consists of Hylleraas functional $J_H$ and the Hartree-Fock energy $E_1$ as in
\begin{align}\nonumber
\mathcal L [\mathbf t, \mathbf \Theta] 
&= E_1[\mathbf \Theta] + J_H[\mathbf t, \mathbf \Theta]\\ \nonumber
&=\sum_i\left<i|h|i\right> + \frac12\sum_{ij} \left<ij||ij\right> 
+ \frac14 \sum_{ijab} \left<ij||ab\right>  t_{ij}^{ab}
+ \frac14 \sum_{ijab} (t_{ij}^{ab})^* \left<ab||ij\right>\\
&\:\:\:\:+ \sum_{ij} P^{(2)}_{ij}F_{ji} + \sum_{ab} P^{(2)}_{ab}F_{ba},
\label{lag}
\end{align}
where the OO and VV MP2 one-particle density matrices (OPDMs) are
\begin{align}
P^{(2)}_{ij} &= -\frac12\sum_{abk}(t_{ik}^{ab})^*t_{jk}^{ab}\\
P^{(2)}_{ab} &= \frac12\sum_{ijc}(t_{ij}^{ac})^*t_{ij}^{bc}.
\end{align}
Assuming pseudocanonical orbitals, the variation in $\mathbf t$ yields
\begin{equation}
t_{ij}^{ab} = -\frac{\left<ab||ij\right> }{\Delta_{ij}^{ab}}, 
\end{equation}
where the denominator is defined as a non-negative quantity,
\begin{equation}
\Delta_{ij}^{ab} = \epsilon_a+\epsilon_b-\epsilon_i-\epsilon_j.
\end{equation}
With the optimal amplitudes, $J_H$ yields the familiar MP2 energy expression,
\begin{equation}
E_\text{MP2} = -\frac14\sum_{ijab} \frac{|\langle ij||ab\rangle|^2}{\Delta_{ij}^{ab}}
\label{eq:emp2}
\end{equation}
We apply the RI approximation\cite{Feyereisen1993,Bernholdt1996}  to the two-electron integrals, 
\begin{equation}
\left<ij|ab\right>
= \sum_{PQ} (ia|P) (P|Q)^{-1} (Q|jb).
\end{equation}
We further define \insertnew{RI fit coefficients, $C_{pq}^P$, for the $|pq\rangle$ charge distribution as:}
\begin{equation}
C^P_{pq}
=
\sum_{pqQ}(P|Q)^{-1} (Q|pq),
\end{equation}
and \insertnew{the 3-center, 2-particle density matrix (TPDM) as}
\begin{equation}
\Gamma^P_{ai}
=
\sum_{jb}
t_{ij}^{ab}C^P_{jb}.
\end{equation}
The TPDM piece of the Hylleraas functional then becomes
\begin{equation}
\frac 12 \sum_{iaP} V_{ia}^P 
\:\Gamma^P_{ai}
+ \text{h.c.},
\end{equation}
where we used
\begin{equation}
V_{ia}^{P} = (ia|P).
\end{equation}
\subsubsection{Orbital Optimization}
The self-consistent field procedure can be described as rotating $N_\text{mo}$ orthonormal vectors until an objective function reaches its stationary point. Thus it is possible to relate two different molecular orbital coefficients with a unitary transformation as in
\begin{equation}
\mathbf C' = \mathbf C \mathbf U,
\end{equation}
where $\mathbf U$ is a unitary (or orthogonal for real variables) matrix. As both the Hartree-Fock energy and the Hylleraas functional in \eqref{lag} are invariant under OO and VV rotations, we consider only the non-redundant OV orbital rotations. We then write the transformation matrix,
\begin{equation}
\mathbf U = \exp[
\mat\Delta_\text{o}\mat\Theta\mat\Delta_\text{v}^\dagger
-\mat\Delta_\text{v}\mat\Theta^\dagger\mat\Delta_\text{o}^\dagger
],
\end{equation}
where $\mat\Delta_\text{o}$ and $\mat\Delta_\text{v}$ are skinny matrices of the dimension $N_\text{mo}\times n_\text{occ}$ and $N_\text{mo}\times n_\text{vir}$, respectively, and they have 1's on the diagonal, and $\mat\Theta$ is a matrix of the dimension $n_\text{occ} \times n_\text{vir}$. It will be useful to expand $\mat U$,
\begin{equation}
\mat U = \mat I + 
\left(
\mat\Delta_\text{o}\mat\Theta\mat\Delta_\text{v}^\dagger
-\mat\Delta_\text{v}\mat\Theta^\dagger\mat\Delta_\text{o}^\dagger
\right)
+ \mathcal O(|\mathbf \Theta|^2)
\end{equation}
Up to first order in $\mat \Theta$, occupied and virtual orbitals transform in the following way:
\begin {align}
C_{\mu i}' = C_{\mu i} - \sum_aC_{\mu a}\Theta^*_{ia}\\
C_{\mu a}' = C_{\mu a} + \sum_iC_{\mu i}\Theta_{ia}
\end{align}
We now consider the variation of the energy, $\delta \mathcal L$, up to first-order in $\mat \Theta$,
\begin{align}\nonumber
\delta \mathcal L &= 
\sum_{ia}
\left(-h_{ai}\Theta_{ia} - h_{ia}\Theta^*_{ia}\right)
-\frac12 \sum_{ija}
\left(
\Braket{ij||aj}\Theta_{ia}^*
+\Braket{ij||ia}\Theta_{ja}^*
+\Braket{aj||ij}\Theta_{ia}
+\Braket{ia||ij}\Theta_{ja}
\right)\\\nonumber
&-\sum_{ijc}P^{(2)}_{ij}\left(
F_{ci}\Theta_{jc} + F_{jc} \Theta_{ic}^*
\right)
-\sum_{ijck} P^{(2)}_{ij}
\left(
\Braket{jc||ik}\Theta_{kc} + \Braket{jk||ic}\Theta_{kc}^*
\right)\\\nonumber
&+\sum_{kab} P^{(2)}_{ab}
\left(
F_{ka}\Theta_{kb}^*
+F_{bk}\Theta_{ka}
\right)
-\sum_{abck}
P^{(2)}_{ab}
\left(
\Braket{bc||ak}\Theta_{kc}
+\Braket{bk||ac}\Theta_{kc}^*
\right)\\
&+
\left[
\frac14
\sum_{ijab}
t_{ij}^{ab}
\left(
-\sum_{c}\left(
\Braket{cj||ab}\Theta_{ic}
+\Braket{ic||ab}\Theta_{jc}
\right)
+\sum_{k}\left(
\Braket{ij||kb}\Theta_{ka}
+\Braket{ij||ak}\Theta_{kb}
\right)
\right)
+\text{h.c.}
\right].
\label{eq:variation}
\end{align}
The first line corresponds to the HF orbital gradient and the rest belongs to the MP2 contribution.
We apply the RI technique we described before to the last line of Eq. \eqref{eq:variation}:
\begin{equation}
\frac12\left(
-V_{ca}^P\Gamma_{ai}^P\Theta_{ic} - V_{cb}^{P}\Gamma_{bj}^P\Theta_{jc}
+V_{ik}^P\Theta_{ka}\Gamma_{ai}^P 
+ V_{jk}^P\Theta_{kb}\Gamma_{bj}^P
\right)
+ \text{h.c.}
\end{equation}
The orbital optimization treats the real and imaginary parts of $\Theta_{ia}$ as separate variables as is done in Wirtinger calculus.\cite{Sorber2012}
\begin{align}
\frac{\delta \mathcal L}{\delta \text{Re(}\Theta_{kc}\text{)}}
&=
\frac{\delta \mathcal L}{\delta \Theta_{kc}}
+\frac{\delta \mathcal L}{\delta \Theta_{kc}^*}\\
\frac{\delta \mathcal L}{\delta \text{Im(}\Theta_{kc}\text{)}}
&=
-i\left(
\frac{\delta \mathcal L}{\delta \Theta_{kc}}
-\frac{\delta \mathcal L}{\delta \Theta_{kc}^*}
\right)
\end{align}
where
\begin{align}\nonumber\label{eq:orbgrad}
\frac{\delta \mathcal L}{\delta \Theta_{kc}} =& 
- F_{ck} - \sum_{i} F_{ci}P_{ik}^{(2)}
+\sum_{a} P_{ca}^{(2)}F_{ak}
-\sum_{ij} P^{(2)}_{ij}
\Braket{jc||ik}
-\sum_{ab}
P^{(2)}_{ab}
\Braket{bc||ak}
\\
&-\sum_{aP} V_{ca}^P\Gamma_{ak}^P
+\sum_{iP} V_{ik}^P\Gamma_{ci}^P
\\
\frac{\delta \mathcal L}{\delta \Theta_{kc}^*} =& 
\left(\frac{\delta \mathcal L}{\delta \Theta_{kc}}\right)^*
\end{align}
In passing we note that the last two terms in Eq. \eqref{eq:orbgrad} are evaluated by the usual mixed Lagrangian technique as used in the nuclear gradient implementation of RI-MP2. \cite{Distasio2007}
\subsection{Regularized OOMP2}
St{\"u}ck and Head-Gordon found a rather disturbing feature of OOMP2 when breaking bonds. \cite{Stuck2013}
The energy denominator $\Delta_{ij}^{ab}$ approaches zero near dissociation limits in the case of restricted orbitals if optimized at the MP2 level. This leads 
to a divergent ROOMP2 solution even when using UOOMP2, as it is variationally preferred. In the perturbation theory literature, this existence of a divergent solution is sometimes referred to as an intruder state problem.


To ameliorate this problem, St{\"u}ck and Head-Gordon employed a simple level-shift scheme to remove the divergent ROOMP2 solution \insertnew{associated with single bond-breaking} and found it somewhat effective. This regularizer will be referred to as a $\delta$-regularizer: $\Delta_{ij}^{ab} \leftarrow \Delta_{ij}^{ab} + \delta$. Some preliminary results on thermochemistry were encouraging with $\delta = 400$ m$E_h$. However, later it was found that the level-shift parameter to restore Coulson-Fischer points for double and triple bond dissociations is too large to give reasonable thermochemistry results. \cite{Razban2017} This sets the main objective of this work. Namely, we are interested in designing a regularizer that can solve the first-order derivative discontinuity and the energy singularity problems while keeping the thermochemistry performance undamaged.

\subsubsection{Design principles of regularizers}
Ideally, one needs a regularizer that damps away a strongly divergent term while keeping physical correlation terms unchanged. 
As an attempt to achieve this goal, we propose multiple classes of orbital energy gap dependent regularizers that remove the singularity problem while (hopefully) damaging thermochemistry results to only a small extent. 

One may understand the MP2 singularity problem based on the following integral transform:
\begin{equation}
E_\text{MP2} = -\frac14\sum_{ijab} \int^{\infty}_0 \text{d}\tau\:
e^{-\tau \Delta_{ij}^{ab}} |\langle ij||ab\rangle|^2
= \frac14\sum_{ijab} \epsilon_{ij}^{ab}
\end{equation}
where
\begin{equation}
\epsilon_{ij}^{ab} =
-\int^{\infty}_0 \text{d}\tau\:
e^{-\tau \Delta_{ij}^{ab}} |\langle ij||ab\rangle|^2 
\end{equation}
This energy expression is derived from a Laplace transformation of the energy expression in Eq. \eqref{eq:emp2} and is a foundation of various linear-scaling MP2 methods.\cite{Almlof1991,Haser1992,Haser1993} When $\Delta_{ij}^{ab} = 0$, the corresponding energy contribution $\epsilon_{ij}^{ab}$ is divergent as the integrand does not decay to zero when $\tau\rightarrow\infty$. 

One may try to regularize $\epsilon_{ij}^{ab}$ by changing $\Delta_{ij}^{ab}$ to $\Delta_{ij}^{ab}+\delta$ with a positive constant $\delta$ so that when $\Delta_{ij}^{ab} = 0$ the integrand decays to zero as $\tau \rightarrow \infty$. This corresponds to the $\delta$-regularizer. Alternatively, one may replace $\Delta_{ij}^{ab}$ in the integrand with a function of $\Delta_{ij}^{ab}$ that does not go to zero as $\Delta_{ij}^{ab} \rightarrow 0$. One such example is 
\begin{equation}
\Delta_{ij}^{ab} + \frac{1}{(\alpha + \Delta_{ij}^{ab})^p}
\end{equation}
where 
$\alpha>0$ and $p$ is a positive integer which can be chosen empirically. This function has a non-zero asymptote for infinitesimal $\Delta_{ij}^{ab}$ 
and becomes $\Delta_{ij}^{ab}$ for large positive values of $\Delta_{ij}^{ab}$. This yields an energy expression,
\begin{equation}
E_\text{MP2}(\alpha,p) =
-\frac14
\sum_{ijab}  \frac{|\langle ij||ab\rangle|^2}
{\Delta_{ij}^{ab} + \frac{1}{(\alpha + \Delta_{ij}^{ab})^p}},
\end{equation}
and we call this class of regularizers $\alpha^p$ regularizers.

Another way to approach this problem is to change the domain of integration for small $\Delta^{ab}_{ij}$ values. The upper limit of the integral should approach $\infty$ for large $\Delta^{ab}_{ij}$ and become a finite value for small $\Delta^{ab}_{ij}$ to remove the singularity. A simple way to achieve this is to have a two-parameter integral upper limit $\sigma ({\Delta_{ij}^{ab}})^p$ where 
$\sigma>0$ and $p$ is a positive integer. We call this regularizer, a $\sigma^p$-regularizer. The regularized energy expression then reads
\begin{equation}
E_\text{MP2}(\sigma,p) 
= 
-\frac14
\sum_{ijab}  \frac{|\langle ij||ab\rangle|^2}
{\Delta_{ij}^{ab}}
\left(
1-e^{-\sigma (\Delta_{ij}^{ab})^p}
\right)
\end{equation}
Interestingly, $p=2$ leads to an energy expression that was derived from the driven similarity renormalization group theory by Evangelista and co-workers.\cite{Evangelista2014}

Lastly, one may modify the two-electron integrals such that the resulting integrand decays to zero when $\Delta_{ij}^{ab}\rightarrow0$. Motivated by the above exponential damping function, we propose to modify $V_{ab}^{ij}$ to
\begin{equation}
W_{ab}^{ij}(\kappa, p) = V_{ab}^{ij}
\left(
1-e^{-\kappa (\Delta_{ij}^{ab})^p}
\right).
\end{equation}
The regularized energy then reads
\begin{equation}
E_\text{MP2}(\kappa,p) =
-\frac14
\sum_{ijab}  \frac{|\langle ij||ab\rangle|^2}
{\Delta_{ij}^{ab}}
\left(
1-e^{-\kappa (\Delta_{ij}^{ab})^p}
\right)^2
\end{equation}
We call this class of regularizers $\kappa^p$ regularizers.

In this work, we shall investigate the $\sigma^1$- and $\kappa^1$-regularizers. These were chosen because one can easily write down a Lagrangian that yields the regularized energy expressions and the orbital gradient is not so complicated.

\subsubsection{$\kappa^1$-Regularizer}
We define the following rank-8 tensor $\mathbf{\Sigma}$ that depends on a single parameter $\beta$,
\begin{equation}
\Sigma_{ijkl}^{abcd} (\beta) = 
(e^{\beta \mathbf{F}^\text{oo}})_{ik}
(e^{\beta \mathbf{F}^\text{oo}})_{jl}
(e^{-\beta \mathbf{F}^\text{vv}})_{ac}
(e^{-\beta \mathbf{F}^\text{vv}})_{bd}
\end{equation}
where $\mathbf{F}^\text{oo}$ and $\mathbf{F}^\text{vv}$ are occupied-occupied and virtual-virtual blocks of Fock matrix, respectively.
If orbitals are pseudocanonical, $\mathbf{\Sigma}$ becomes much sparser:
\begin{equation}
\Sigma_{ijkl}^{abcd} (\beta)
 = 
 e^{-\beta\Delta_{ij}^{ab}} \delta_{ik}\delta_{jl}\delta_{ac}\delta_{bd}
\end{equation}
We write the regularized OOMP2 Lagrangian modifying the two-electron integrals in Eq. \eqref{eq:hylleraas},
\begin{equation}
\mathbf{t}^\dagger \mathbf{V} \rightarrow \mathbf{t}^\dagger (\mathbf{1}-\mathbf{\Sigma}(\kappa)) \mathbf{V}
\equiv
\mathbf{t}^\dagger \mathbf{W}(\kappa)
\end{equation}
where
the damped integral $\mathbf{W}$ is defined as
\begin{equation}
\mathbf{W}(\kappa) = 
(\mathbf{1}-\mathbf{\Sigma}(\kappa)) \mathbf{V}
\label{eq:damped}
\end{equation}
Using this, the modified Lagrangian reads
\begin{equation}
{\mathcal L}[{\mathbf t}, \mathbf \Theta] = 
\mathcal L_0 [{\mathbf t}, \mathbf \Theta]
-\mathbf{V}^\dagger\mathbf{\Sigma}(\kappa) {\mathbf t}
-{\mathbf t}^\dagger\mathbf{\Sigma}(\kappa) \mathbf{V}
\label{eq:mlag}.
\end{equation}
This leads to modified amplitudes,
\begin{equation}
{t}_{ij}^{ab} = -\frac{\langle ab||ij\rangle}{\Delta_{ij}^{ab}}\left(1-e^{-\kappa\Delta_{ij}^{ab}}\right)
\end{equation}
In the limit of $\Delta_{ij}^{ab}\rightarrow 0$, ${t}_{ij}^{ab} \rightarrow -\kappa \langle ab||ij\rangle$ as opposed to $\infty$.
The regularized MP2 energy from the modified Lagrangian follows
\begin{equation}
{E}_\text{MP2}(\kappa) = - \frac14 \sum_{ijab} \frac{|\langle ij||ab\rangle|^2}{\Delta_{ij}^{ab}} \left(1-e^{-\kappa\Delta_{ij}^{ab}}\right)^2,
\label{eq:energy2}
\end{equation}
which is the $\kappa^1$-OOMP2 energy.
We note that $\Delta_{ij}^{ab} \rightarrow 0$ does not contribute to the energy. Obviously, the large $\kappa$ limit recovers the unregularized energy expression.

The orbital gradient is simply the sum of Eq. \eqref{eq:orbgrad} (where $\mathbf P^\text{(2)}$ and $\mathbf \Gamma$ are computed with modified amplitudes) and the contribution from the two additional terms in Eq. \eqref{eq:mlag}.
In the pseudocanonical basis,
\begin{equation}
\mathbf{V}^\dagger\mathbf{\Sigma}(\kappa) {\mathbf t}
=
\frac14
\sum_{ijab}
 e^{-\kappa\Delta_{ij}^{ab}}
 \langle ij||ab\rangle
t_{ij}^{ab}
\end{equation}
Differentiating $\langle ij||ab\rangle$ was already explained before, so we focus on the derivative of $ e^{\kappa\Delta_{ij}^{ab}}$.
We have
\begin{equation}
\frac{\partial e^{\kappa {\mathbf{F}}^\text{oo}}}
{\partial \Theta_{kc}}
=\kappa
\int_{0}^1 \text{d}\tau\:\:
e^{(1-\tau)\kappa{\mathbf{F}}^\text{oo}}
\frac{\partial \mathbf{F}^\text{oo}}
{\partial \Theta_{kc}}
e^{\tau\kappa{\mathbf{F}}^\text{oo}}
\end{equation}
 and
\begin{equation}
\frac{\partial F_{ij}}
{\partial \Theta_{kc}}
=
-F_{cj}\delta_{ik} - \langle ic || jk \rangle
\end{equation}
Similarly,
\begin{equation}
\frac{\partial e^{-\kappa {\mathbf{F}}^\text{vv}}}
{\partial \Theta_{kc}}
= -\kappa
\int_{0}^1 \text{d}\tau\:\:
e^{-(1-\tau)\kappa{\mathbf{F}}^\text{vv}}
\frac{\partial \mathbf{F}^\text{vv}}
{\partial \Theta_{kc}}
e^{-\tau\kappa{\mathbf{F}}^\text{vv}}
\end{equation}
and
\begin{equation}
\frac{\partial F_{ab}}
{\partial \Theta_{kc}}
=
F_{ak}\delta_{bc} - \langle ac || bk \rangle
\end{equation}
Generally, one needs to perform an one-dimensional quadrature to compute this contribution to the orbital gradient.
In the pseudocanonical basis, the extra contribution to the orbital gradient is given as
\begin{align}\nonumber
&\frac{\partial(\mathbf{V}^\dagger\mathbf{\Sigma}(\kappa) {\mathbf t} + \text{h.c.})}
{\partial \Theta_{kc}}
=
-\sum_{aP}
e^{-\kappa \Delta_k^a}V^P_{ca}\tilde{\Gamma}^P_{ak}
+\sum_{iP}
e^{-\kappa \Delta_i^c}
\tilde{\Gamma}^P_{ci}
B^P_{ik}
\\\nonumber
&-\kappa 
\int_{0}^1 \text{d}\tau\:\:
\left(
\tilde{F}_{ck}(\tau) e^{(1-\tau)\kappa\epsilon_k}
+
\tilde{\tilde{F}}_{ck}(\tau) e^{-\tau\kappa\epsilon_c}
\right)\\
&-
\kappa 
\int_{0}^1 \text{d}\tau\:\:
\left(\sum_{\mu\nu}
\left[
\left(X_{\nu\mu}(\tau) - Y_{\nu\mu}(\tau)\right)
\langle
\mu c||\nu k
\rangle
\right]
\right)
\label{eq:quadrature}
\end{align}
where we define
\begin{equation}
\Delta_p^q = \epsilon_q - \epsilon_p,
\end{equation}
\begin{equation}
\tilde{\Gamma}_{ai}^P = \sum_{jb} e^{-\kappa\Delta_j^b}{t}_{ij}^{ab} B_{jb}^P,
\label{eq:gammap}
\end{equation}
\begin{equation}
\tilde{F}_{ck}(\tau) = \sum_l F_{cl}e^{\tau\kappa\epsilon_l}(\omega_{lk}^*+\omega_{kl}),
\end{equation}
\begin{equation}
\tilde{\tilde{F}}_{ck}(\tau) = \sum_a (\omega_{ca}+\omega_{ac}^*)e^{-(1-\tau)\kappa\epsilon_a}F_{ak},
\end{equation}
\begin{equation}
X_{\nu\mu}(\tau)
=
\sum_{ij}
e^{\tau\kappa\epsilon_j}
C_{\nu j}
(\omega_{ji}^* + \omega_{ij})
C_{\mu i}^*
e^{(1-\tau)\kappa\epsilon_i}
\end{equation}
\begin{equation}
Y_{\nu\mu}(\tau)
=
\sum_{ab}
e^{-\tau\kappa\epsilon_b}
C_{\nu b}
(\omega_{ba} + \omega_{ab}^*)
C_{\mu a}^*
e^{-(1-\tau)\kappa\epsilon_a},
\end{equation}
\begin{equation}
\omega_{lk} = \sum_{aP} e^{-\kappa\epsilon_a}
V_{la}^P 
\tilde{\Gamma}_{ak}^P,
\end{equation}
and
\begin{equation}
\omega_{ac} = \sum_{iP}  e^{\kappa\epsilon_i}
\tilde{\Gamma}_{ai}^PV_{ic}^P.
\end{equation}
Those extra terms can be readily implemented to any existing OOMP2 programs and there is only a mild increase in the computational cost.
The only additional $\mathcal O(N^5)$ step is the formation of $\tilde{\mathbf{\Gamma}}$ in Eq. \eqref{eq:gammap} and this can be done at the same time as forming $\mathbf{\Gamma}$.

\subsubsection{$\sigma^1$-Regularizer}

The $\sigma^1$-regularizer can be derived from a Hylleraas functional with a set of auxiliary amplitudes $\mathbf s$. We write the new Hylleraas functional in the following way:
\begin{equation}
J_H[\mathbf t,\mathbf s,\mathbf \Theta] = 
\frac12\mathbf{s}^\dagger \Delta \mathbf{t} 
+ \frac12\mathbf{s}^\dagger \mathbf{W} (\sigma)
+ \frac12\mathbf{t}^\dagger \mathbf{V}
+ \text{h.c.}
\label{eq:hylleraas3}
\end{equation}
where $\mathbf W(\sigma)$ is the damped integral defined in Eq. \eqref{eq:damped}. 
The modified Hylleraas functional is now a functional of $\mathbf t$, $\mathbf s$, and $\mathbf \Theta$.
Stationary conditions on $\mathbf s^\dagger$ and $\mathbf t^\dagger$ yields
\begin{align}\label{eq:amp1}
\mathbf s &= -\mathbf \Delta ^{-1} \mathbf V, \\
\label{eq:amp2}
\mathbf t &= -\mathbf \Delta ^{-1} \mathbf W.
\end{align}

Plugging Eq. \eqref{eq:amp1} and Eq. \eqref{eq:amp2} to Eq. \label{eq:hylleraas3} leads to the following energy expression:
\begin{equation}
{E}_\text{MP2}(\sigma) = -\frac14 \sum_{ijab} \frac{|\langle ij||ab\rangle|^2}{\Delta_{ij}^{ab}} \left(1-e^{-\sigma\Delta_{ij}^{ab}}\right),
\label{eq:energy3}
\end{equation}
which is the $\sigma^1$-OOMP2 energy expression.
Unlike Eq. \eqref{eq:energy2}, Eq. \eqref{eq:energy3} has non-zero contributions from small $\Delta_{ij}^{ab}$ as the limit yields
$-\sigma |\langle ij||ab\rangle|^2$. The orbital gradient is more or less the same as that of $\kappa^1$-OOMP2.
$\sigma^1$-OOMP2 can also be implemented with a moderate increase in the computational cost. 
\section{Computational Details}
All the calculations presented below are carried out by the development version of Q-Chem. \cite{Shao2015} The self-consistent field calculations are based on Q-Chem's new object-oriented SCF library, \texttt{libgscf} and the relevant MP2 components are implemented through Q-Chem's new MP2 library, \texttt{libgmbpt}. All those implementations are already at the production level and OpenMP parallelized. All the correlated wave function calculations presented here were performed with all electrons correlated and all virtual orbitals included unless specified otherwise.

The quadrature evaluation in Eq. \eqref{eq:quadrature} was performed using the standard Gauss-Legendre quadrature. The accuracy of the quadrature for a given quadrature order depends on the orbital energies and thus it is system-dependent. For systems presented below, 20 quadrature points were found to be sufficient. The precise assessment of the accuracy of the quadrature will be an interesting subject for the future study.
\section{Results and Discussion}

For the sake of simplicity, we will refer $\kappa^{1}$ and $\sigma^{1}$-regularizers to as $\kappa$- and $\sigma$-regularizers, respectively.

\subsection{ROOMP2 to UOOMP2 Stability Analysis}
As noted before, we would like UOOMP2 to spontaneously spin-polarize to reach the proper dissociation limits. Without regularization, it is quite common to observe that UOOMP2 stays on an R solution and never spontaneously polarizes even though there is a more stable U solution that dissociates correctly. This may not be a serious problem if stability analysis can detect those more stable polarized solutions. However, in most cases, there exists a barrier between R and U solutions so that both solutions are stable up to the quadratic stability analysis.

We revisit and assess the new regularizers on the bond-breaking of \ce{H2} (single-bond), \ce{C2H6} (single-bond), \ce{C2H4} (double-bond), and \ce{C2H2} (triple-bond). 
We present the results for unregularized, $\kappa$-, and $\sigma$- OOMP2. Interested readers are referred to ref. \citenum{Razban2017} for the $\delta$-OOMP2 result. 
The main objective of this section is to find out whether there is a reasonably weak single parameter $\kappa$ or $\sigma$ that recovers the Coulson-Fischer points\cite{Coulson1949} (i.e. the geometries at which spontaneous, continuous symmetry breaking start to occur) for all four cases.
All the results are obtained with the cc-pVDZ basis set \cite{Dunning1989} along with its auxiliary basis set. \cite{Weigend2002}
\revision{The diagonalization of the R to U stability Hessian was performed iteratively with the Davidson solver \cite{Davidson1975} based on the finite difference matrix-vector product technique developed in ref. \citenum{Sharada2015}. This technique utilizes the analytic orbital gradient and does not require the implementation of the analytic orbital Hessian.}

\begin{figure}[h!]
\includegraphics[scale=0.5]{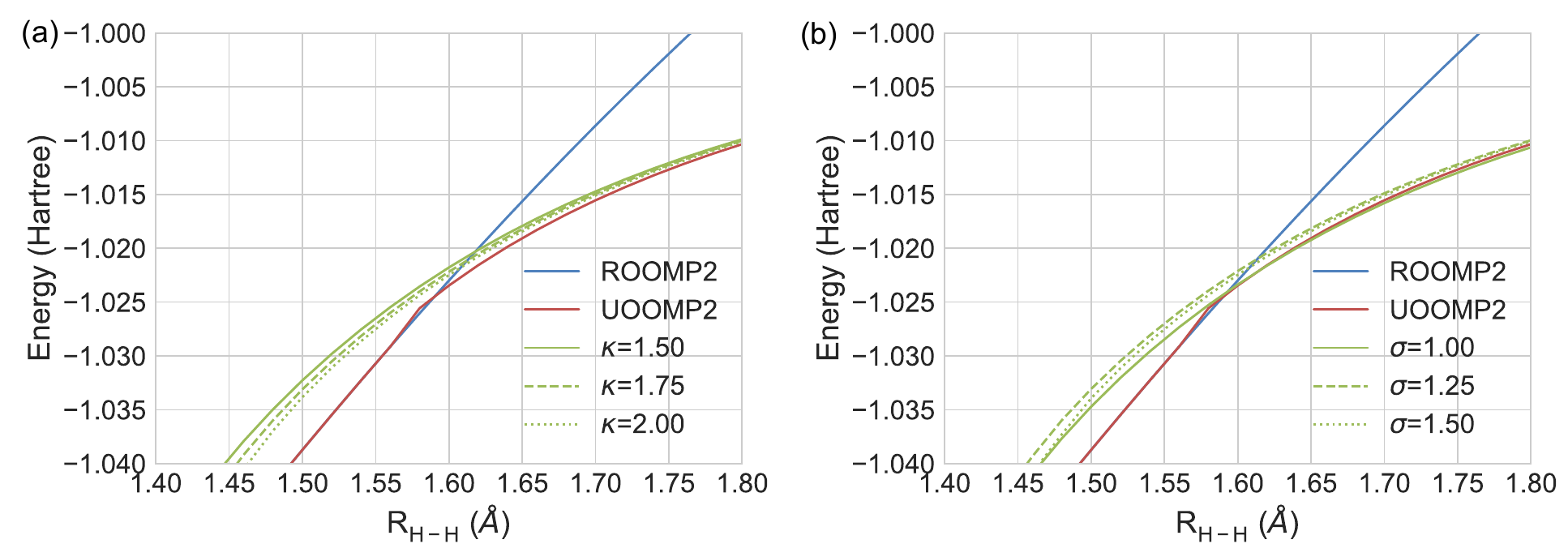}
\caption{\label{fig:h2}
The potential energy curve of \ce{H2} within the cc-pVDZ basis set.
All the regularized OOMP2 are performed with spin-unrestriction. (a) $\kappa$-OOMP2 and (b) $\sigma$-OOMP2.
}
\end{figure}

In Figure \ref{fig:h2}, we present the potential energy curve (PEC) of the \ce{H2} dissociation. It is clear that there is a lower U solution starting from 1.6 \AA \:and this is the solution that dissociates properly. However, there is a slight first-order discontinuity between 1.58 \AA \:and 1.60 \AA. This was previously noted by one of us in ref. \citenum{Razban2017}. On the other hand, both $\kappa$- and $\sigma$-UOOMP2 exhibit smooth curves and dissociate properly. We picked the range of $\kappa$ and $\sigma$ based on the absolute energies that yield 1-2 m$E_h$ higher than the unregularized one at the equilibrium geometry. The precise determination of those values will be discussed in the next section. 

\begin{figure}[h!]
\includegraphics[scale=0.5]{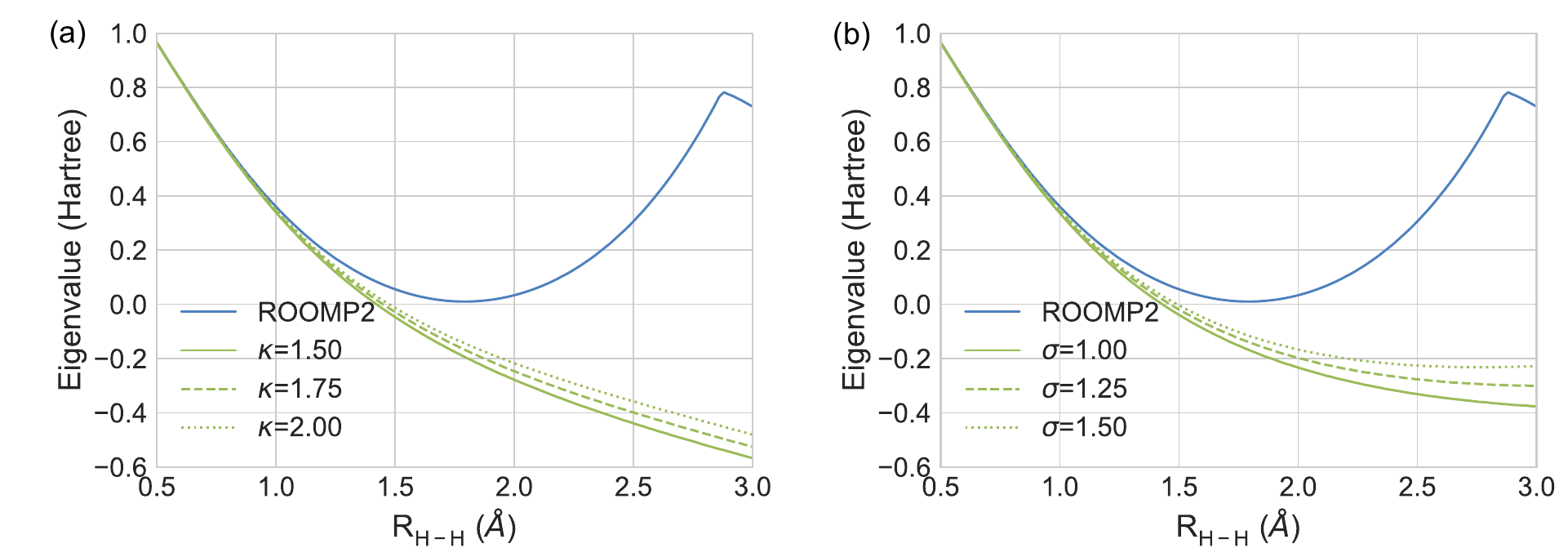}
\caption{\label{fig:h2stab}
The R to U Hessian lowest eigenvalues of \ce{H2} within the cc-pVDZ basis set. (a) $\kappa$-OOMP2 and (b) $\sigma$-OOMP2.
}
\end{figure}
The continuous transition from R to U in regularized OOMP2 can be understood based on the R to U Hessian lowest eigenvalues as illustrated in Figure \ref{fig:h2stab}. The unregularized ROOMP2 exhibits no R to U instability at every given bond distance. In other words, there is no solution that smoothly and barrierlessly connects this R solution to the lower U solution in UOOMP2. This is the source of the first-order discontinuity of ROOMP2 in Fig. \ref{fig:h2}. 
In contrast to the unregularized ROOMP2 result, both $\kappa$- and $\sigma$-ROOMP2 exhibit negative eigenvalues after the critical bond length around 1.5 \AA. This results into a continuous and smooth transition from the R solution to the U solution as we stretch the bond.

\begin{figure}[h!]
\includegraphics[scale=0.5]{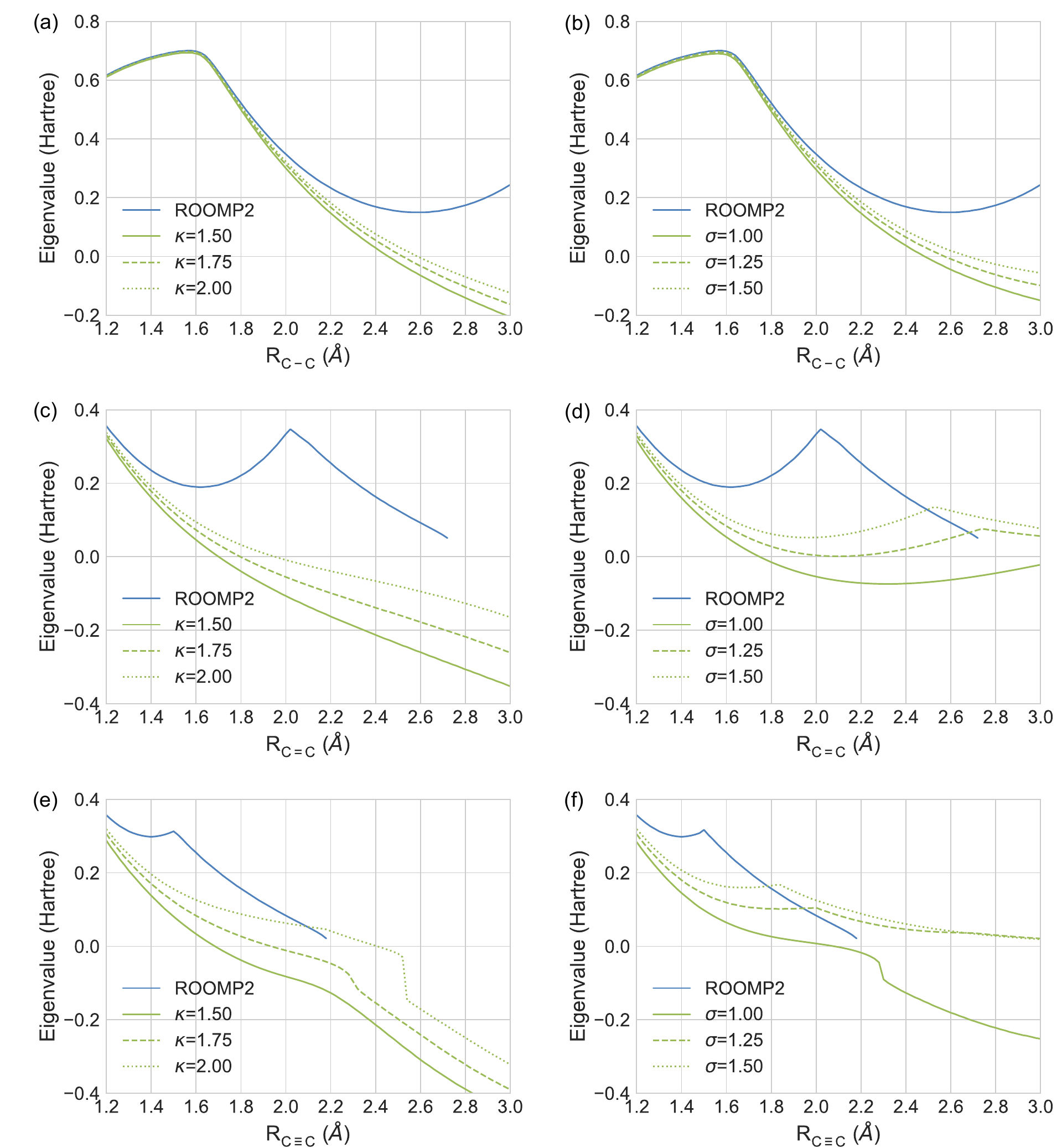}
\caption{\label{fig:ethanestab}
The R to U Hessian lowest eigenvalues of ethane, ethene and ethyne within the cc-pVDZ basis set.
(a) Ethane with $\kappa$-ROOMP2.
(b) Ethane with $\sigma$-ROOMP2.
(c) Ethene with $\kappa$-ROOMP2.
(d) Ethene with $\sigma$-ROOMP2.
(e) Ethyne with $\kappa$-ROOMP2. 
(f) Ethyne with $\sigma$-ROOMP2. 
Note that discontinuities in the plots indicate discontinuous transitions in the corresponding ROOMP2 curves.
ROOMP2 does not converge after 2.72 \AA \:\:for ethene and 2.18 \AA \:\:for ethyne.
}
\end{figure}
We repeat this analysis for ethane, ethene, and ethyne. The corresponding R to U Hessian lowest eigenvalues are plotted in Figure. \ref{fig:ethanestab}.
ROOMP2 shows numerical instabilities in the case of ethene and ethyne whereas both of the regularized ones converge properly.
In all cases, the $\kappa$ and $\sigma$ regularizers show clear differences. 
Namely, the $\sigma$ regularizer shows clearly slower appearance of the negative roots compared to the $\kappa$ case. Furthermore, the eigenvalue of the $\sigma$ regularizer tends to turn around after some distance; this is not desirable as there can be a discontinuous transition between R and U solutions depending on where we start. 
\insertnew{Furthermore, the bond length, at which the Coulson-Fischer point is located, is longer in ethene than in ethyne in the case of $\sigma$-OOMP2.}
Given these results, we would prefer the $\kappa$ regularizer over the $\sigma$ regularizer. However, a more detailed assessment is necessary to make a general recommendation.

\subsection{Training the Regularization Parameter and Its Validation}
\subsubsection{The W4-11 set}
Though the investigation of the stability Hessian eigenvalues was informative, it is not sufficient to suggest a value for the regularization parameter to be used for general chemical applications.
Training the regularization parameter on a minimal training set and validating on other test sets will provide a sensible preliminary value. 

We chose the W4-11 set developed by Martin and co-workers for the training set. \cite{Karton2011} The W4-11 set has played a crucial role in density functional development. 
\revision{The W4-11 set consists of the following subsets: 140 total atomization energies (TAE140), 99 bond dissociation energies (BDE99), 707 heavy-atom transfer (HAT707), 20 isomerization (ISOMER20), and 13 nucleophilic substitution reaction (SN13).}
MP2 and OOMP2 do not perform very well on this set (i.e. root-mean-square-deviation (RMSD) of 15.10 kcal/mol and 11.08 kcal/mol within the aug-cc-pVTZ basis,\cite{Dunning1989} respectively). Therefore, it is sensible to choose it as a training set as an attempt to improve upon both MP2 and OOMP2.

The training was done within the aug-cc-pVTZ (aVTZ) basis set \cite{Dunning1989} along with the corresponding auxiliary basis set.\cite{Weigend2002} All the calculations are performed with the geometric direct minimization (GDM) algorithm \cite{Voorhis2002} and a stable UHF solution. OO methods used a stable UHF solution as a guess. We also performed additional training at the complete basis set (CBS) limit using the aVTZ and aug-cc-pVQZ (aVQZ) \cite{Dunning1989} extrapolation (i.e. the TQ extrapolation). \cite{Helgaker1998} 

\begin{figure}[h!]
\includegraphics[scale=0.5]{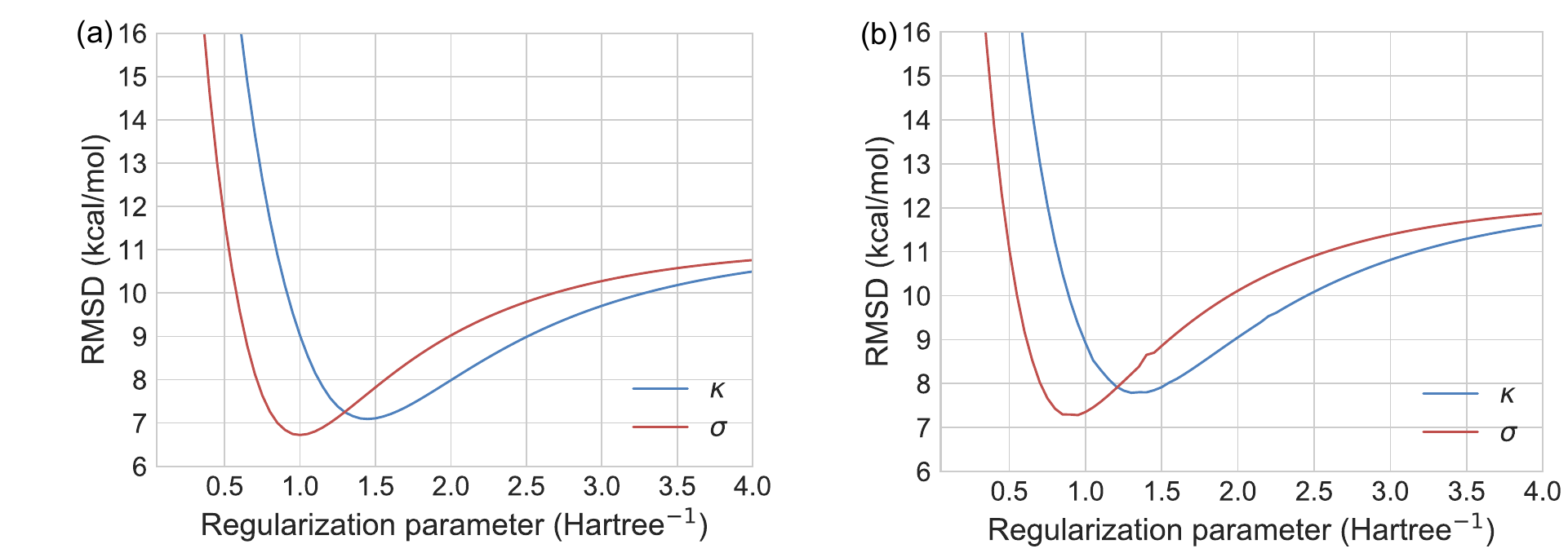}
\caption{\label{fig:w411rmsd}
The W4-11 set RMSD (kcal/mol)
as a function of regularization parameter $\kappa$ and $\sigma$. (a) aVTZ and (b) TQ extrapolated results.
The optimal values are $\kappa =$1.45 $E_h^{-1}$ and  $\sigma =$ 1.0 $E_h^{-1}$ for aVTZ and $\kappa =$ 1.40 $E_h^{-1}$ and  $\sigma =$ 0.95 $E_h^{-1}$ for the TQ extrapolated case.
}
\end{figure}

Initially, we optimized only one non-linear parameter ($\kappa$ or $\sigma$) by scanning over [0.05, 4.0]. The optimal values were $\kappa=1.45$ with the RMSD of 7.09 kcal/mol and $\sigma=1.00$ with the RMSD of 6.72 kcal/mol as shown in Fig. \ref{fig:w411rmsd} (a). 
The optimal values change slightly for the TQ extrapolated results as in Fig. \ref{fig:w411rmsd} (b). We have $\kappa=1.40$ and $\sigma=0.95$ with the RMSD of 7.80 kcal/mol and 7.28 kcal/mol, respectively.
These values are also enough to restore the Coulson-Fischer points in the systems studied before. We recommend those values for general applications.

We also developed a scaled correlation energy variant by adding a linear parameter that scales the overall correlation energy to improve the thermochemistry performance. This is achieved by optimizing a linear parameter $c$ in
\begin{equation}
E_{tot} = E_{HF} + c E_{MP2}
\end{equation}
in addition to the non-linear parameter ($\kappa$ or $\sigma$) for the regularizer. $c$ can be optimized by a means of least-squares-fit. Since changing $c$ alters orbitals, one needs to accomplish self-consistency when optimizing $c$. We found three self-consistency cycles enough to converge $c$ and present only the final result. 
For preliminary results, we simply trained this over the TAE-140 set where unregularized OOMP2 performs much worse (an RMSD of 17.48 kcal/mol) than it does in the other subsets of the W4-11 set. 
\begin{figure}[h!]
\includegraphics[scale=0.5]{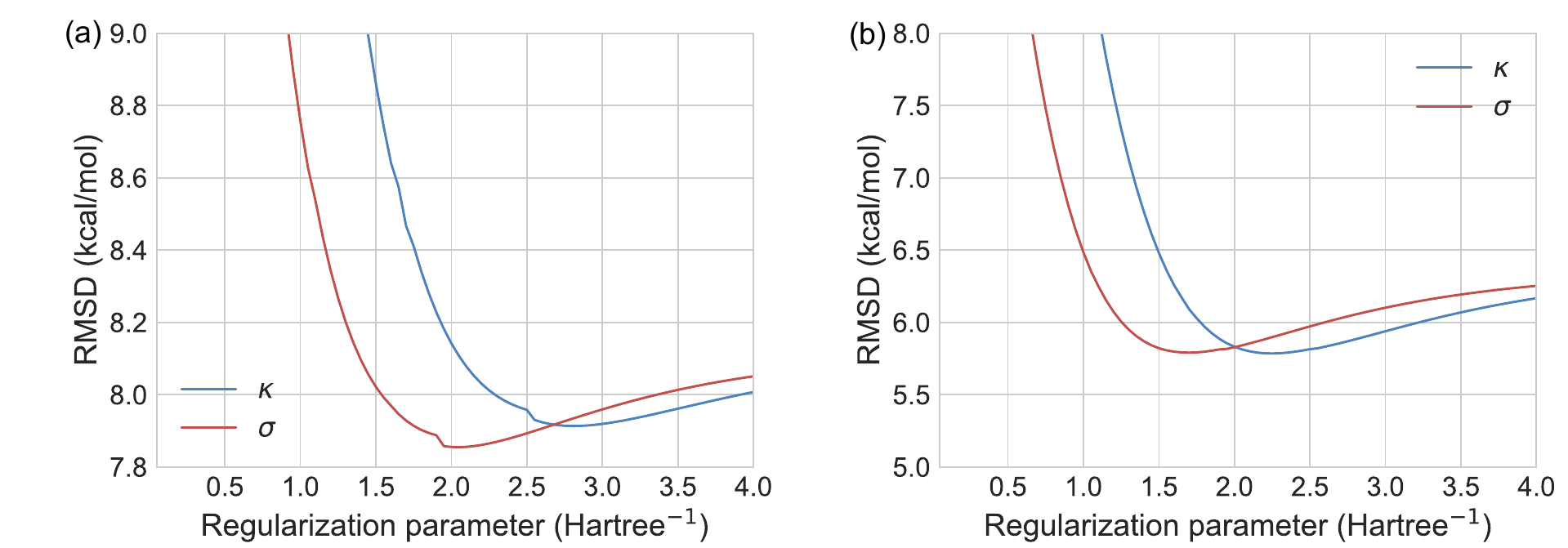}
\caption{\label{fig:scsrmsd}
(a) The TAE140 set RMSD and (b) the W4-11 set RMSD
as a function of regularization parameter $\kappa$ and $\sigma$ where the scaling parameter $c$ for each data point is optimal within the TAE-140 set.
Discontinuities are caused by the appearance of different orbital solutions in the MR16 subset of the TAE140 set.
The basis set used was aVTZ.
}
\end{figure}

In Figure \ref{fig:scsrmsd}, we present the RMSD values of the TAE140 set and the entire W4-11 set as a function of the regularization strength. All the RMSD values are obtained with the scaling factor optimized for the TAE-140 set. In Figure \ref{fig:scsrmsd} (a), we have optimal values of $\kappa=2.80$ and $\sigma=2.05$. These values are not enough to restore the Coulson-Fischer points. Validating those values over the entire W4-11 set as in Figure \ref{fig:scsrmsd} (b) shows that those values are not optimal on the entire W4-11 set. The optimal values are $\kappa=2.25$ and $\sigma=1.70$ and those are not enough to restore the Coulson-Fischer points.

For general applications, we recommend the values of $\kappa=1.50$ along with $c = 0.955$ and $\sigma=1.00$ along with $c = 0.973$.
Both yield the RMSD of 6.48 kcal/mol over the entire W4-11 set. We refer each model to as $\kappa$-S-OOMP2 and $\sigma$-S-OOMP2, respectively. Although the scaling factor is not optimal anymore, those values are still an improvement over their unscaled variants on the W4-11 set.
The values are slightly smaller than 1.00 because orbital optimization can often overcorrelate.

The TQ extrapolated results are qualitatively similar and we only mention the final results here. We have $\kappa=1.50$ along with $c = 0.931$ with the RMSD of 6.61 kcal/mol and $\sigma=1.00$ along with $c = 0.949$ with the RMSD of 6.63 kcal/mol. These are more or less identical results to those of aVTZ with a smaller scaling factor $c$. Therefore, for the rest of this paper, we will use parameters optimized for the aVTZ basis set.
\begin{table}
\includegraphics[scale=0.54]{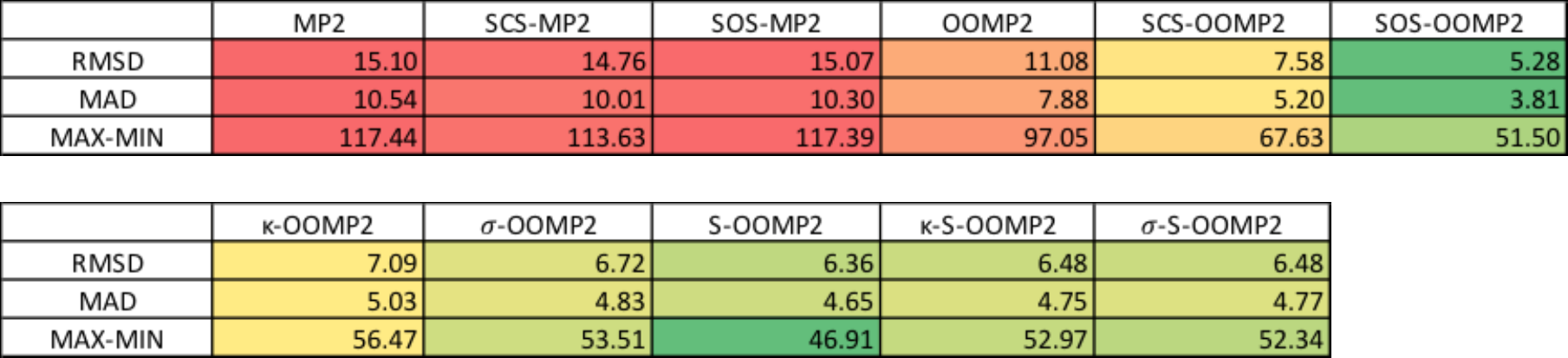}
\caption{\label{tab:w411rmsd}
The W4-11 set RMSD (kcal/mol)
of MP2, OOMP2, and their variants. MAD stands for Mean Absolute Deviation and MAX-MIN stands for the difference between maximum and minimum. The colors represent the relative performance of each method; red means the worst and green means the best among the methods presented. For SOS-MP2 and SOS-OOMP2, the Laplace transformation trick is used. All the calculations were performed with the aVTZ basis set.
}
\end{table}

We summarize the resulting regularized OOMP2 methods in Table \ref{tab:w411rmsd} along with MP2, OOMP2 and other variants of them. All of them were performed with the aVTZ basis set. In particular, we compare the regularized OOMP2 methods with SCS-\cite{Grimme2003} and SOS-MP2 \cite{Jung2004} and their OOMP2 variants.\cite{Lochan2007,Neese2009} For SCS-MP2 and SCS-OOMP2, $c_{ss} = 1/3$ and $c_{os} = 6/5$ are used.\cite{Grimme2003,Neese2009} $c_{os} = 1.3$ is used for SOS-MP2\cite{Jung2004}and $c_{os} = 1.2$ is used for SOS-OOMP2.\cite{Lochan2007} \insertnew{For comparison, we developed a single scaling parameter OOMP2 (S-OOMP2) where the parameter was fitted to the TAE140 set. The optimal scaling parameter is $c=0.90$.}

In Table \ref{tab:w411rmsd}, MP2 performs the worst among the methods examined in this work and SCS-MP2 and SOS-MP2 provide only a small improvement over MP2. OOMP2 improves about 4 kcal/mol in the RMSD compared to MP2. SCS-OOMP2 shows roughly a factor of 2 improvement over SCS-MP2. SOS-OOMP2 performs the best among the methods presented and shows a 2.5 times smaller RMSD than that of SOS-MP2. RMSD, MAD, and MAX-MIN show the same trend. 

The unscaled regularized OOMP2 methods, $\kappa$-OOMP2 and $\sigma$-OOMP2, both provide improved energetics compared to the unregularized one. They are comparable to SCS-OOMP2 and SOS-OOMP2. However, it should be noted that the regularization parameters are optimized for the W4-11 set. It is not so surprising that $\kappa$-OOMP2 and $\sigma$-OOMP2 perform relatively well.

S-OOMP2 performs better than those unscaled regularized OOMP2. Given that S-OOMP2 was trained over only the TAE140 set, this is an interesting outcome. Adding regularizers to S-OOMP2 provides no improvement and it makes the performance little worse. However, the regularization is necessary to restore the Coulson-Fischer points for molecules studied in the previous section. We also note that $\kappa$-S-OOMP2 and $\sigma$-S-OOMP2 exhibit a more or less identical performance in terms of RMSD, MAD, and MAX-MIN.

\subsubsection{The RSE43 set}
We also validate the regularized OOMP2 methods on the RSE43 set \cite{Zipse2006} where unregularized OOMP2 performs nearly perfectly.\cite{Neese2009} The RSE43 has total 43 radical stabilization energies and all of them are energies of a reaction where a methyl radical abstracts a hydrogen from a hydrocarbon. The original RSE43 set reference values were not considered of very high quality\cite{Zipse2006} and thus we use the updated reference data based on the work by Grimme and co-workers.\cite{Goerigk2017} 

We compare MP2, OOMP2, their variants, and three combinatorially designed density functionals ($\omega$B97X-V, $\omega$B97M-V, and B97M-V) developed in our group. 
$\omega$B97X-V is a range-separated generalized gradient approximation (GGA) hybrid functional with the VV10 dispersion tail and $\omega$B97M-V is a range-separated meta GGA with the VV10 dispersion tail. B97M-V is a meta GGA pure functional with the VV10 dispersion tail. All DFT calculations are performed with the aVQZ basis set \cite{Dunning1989} and all the MP2 and OOMP2 calculations are done with the aVTZ basis set.\cite{Dunning1989} MP2 calculations are performed with a stable UHF solution and OOMP2 calculations started from a stable UHF solution.

\begin{table}
\includegraphics[scale=0.53]{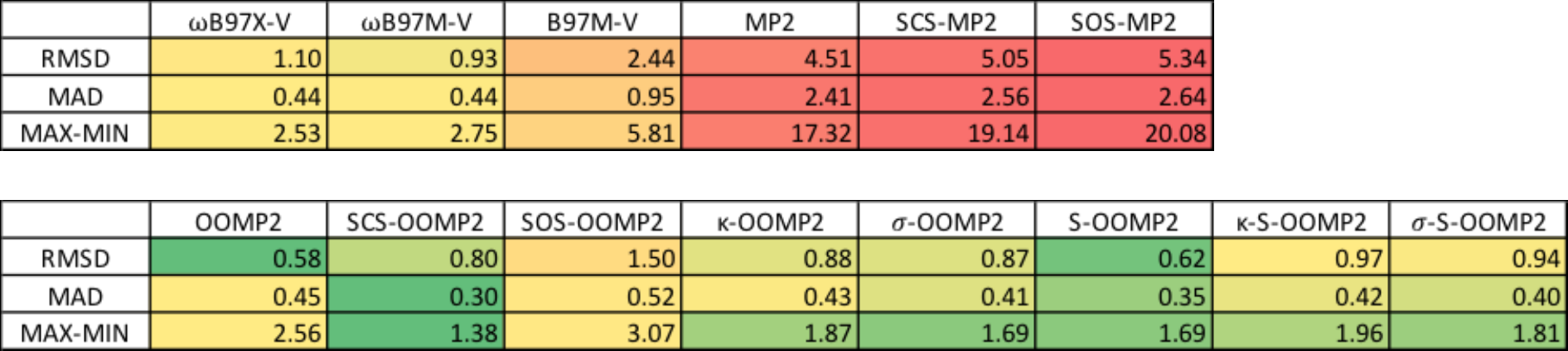}
\caption{\label{tab:rse43}
The RSE43 set RMSD (kcal/mol)
of $\omega$B97X-V, $\omega$B97M-V, and B97M-V, MP2, OOMP2, and their variants. MAD stands for Mean Absolute Deviation and MAX-MIN stands for the difference between maximum and minimum. The colors represent the relative performance of each method; red means the worst and green means the best among the methods presented.
}
\end{table}

The RSE43 set RMSD values are presented in Table \ref{tab:rse43}. DFT functionals outperform MP2, SCS-MP2, and SOS-MP2. The poor quality of those MP2 methods is likely because of the artificial spin symmetry breaking at the HF level. 
DFT functionals are in general less prone to the artificial spin symmetry breaking problem. 
$\omega$B97X-V and $\omega$B97M-V exhibit nearly identical results and B97M-V is roughly twice worse than those two in terms of RMSD, MAD, and MAX-MIN. 

Orbital optimization generally improves the energetics here. 
SOS-OOMP2 does not outperform OOMP2 and SCS-OOMP2 in this case. S-OOMP2 is comparable to OOMP2 and SCS-OOMP2. 
$\kappa$-OOMP2 and $\sigma$-OOMP2 along with their scaled variants are comparable to the unregularized ones. Adding linear parameters on top of those to scale the MP2 correlation energy does not alter the results significantly.
Since the regularization damps out the absolute MP2 energy quite significantly, this is a non-trivial and exciting result.

\subsubsection{The TA13 set}

We further test our new methods on the TA13 set.\cite{Tentscher2013} This dataset involves 13 radical-closed-shell non-bonded interaction energies. We used the aVTZ basis set \cite{Dunning1989} and counterpoise corrections to mitigate basis set superposition error (BSSE). Examining spin-contamination, one data point, \ce{CO+}, was found to be an outlier. The UHF $\langle S^2\rangle$ for \ce{CO+} is 0.98, which deviates significantly from its ideal value of 0.75. With OOMP2, the zeroth order $\langle S^2\rangle$ is 0.76, which is quite close to the ideal value. The same is observed in the case of regularized OOMP2.

In Table \ref{tab:ta13big}, we present the interaction energy errors for each data point of MP2, OOMP2, and their variants. 
Going from MP2 to OOMP2, there are several noticeable changes. The problematic \ce{HF-CO+} interaction is handled much better with OO.
Another problematic case in MP2 is the \ce{H2O-F} interaction and this is also improved with OO. 
\replacewith{The \ce{FH-NH2} case is interesting as none of the methods presented seem to describe it in a satisfactory way. }{}
Overall, without OO, SCS-MP2 and SOS-MP2 are not any better than MP2. Comparing SCS- and SOS-OOMP2 with OOMP2, scaling does not help improve the energetics of OOMP2 and in fact tends to make it worse. $\kappa$- and $\sigma$-OOMP2 perform more or less the same, as do their scaled variants. Regularization \replacewith{tends to make the performance of OOMP2 a little worse but}{keeps the performance of OOMP2 unchanged.} The regularized ones perform better than simple scaled OOMP2 methods (i.e. SCS-OOMP2, SOS-OOMP2, and S-OOMP2). 
\revision{It is interesting that the one-parameter model, S-OOMP2, performs better than the more widely used two-parameter models, SCS-OOMP2 and SOS-OOMP2.}

\begin{sidewaystable}
\centering
  \begin{tabular}{c|r|r|r|r|r|r|r|r|r|r|r}\hline
Complexes& MP2& \shortstack{SCS-\\MP2}& \shortstack{SOS-\\MP2}& OOMP2& \shortstack{SCS-\\OOMP2}& \shortstack{SOS-\\OOMP2}& 
\shortstack{$\kappa$-\\OOMP2}& \shortstack{$\sigma$-\\OOMP2}& 
\shortstack{S-\\OOMP2}& \shortstack{$\kappa$-S-\\OOMP2}& \shortstack{$\sigma$-S-\\OOMP2}\\\hline
\ce{H2O}-\ce{Al}&    1.42& 2.77& 3.45& 0.76& 2.18& 3.00& 1.00& 1.06& 1.13& 1.13& 1.15\\
\ce{H2O}-\ce{Be+}&   2.38& 2.84& 3.08& 2.67& 3.09& 3.09& 1.84& 1.71& 2.45& 1.85& 1.68\\
\ce{H2O}-\ce{Br}&    1.11& 2.01& 2.45& 0.23& 1.39& 2.22& 0.95& 1.03& 0.79& 1.11& 1.15\\
\ce{HOH}-\ce{CH3}&   0.14& 0.49& 0.66& -0.01& 0.39& 0.67& 0.08& 0.12& 0.16& 0.14& 0.16\\
\ce{H2O}-\ce{Cl}&    1.32& 2.20& 2.64& 0.13& 1.37& 2.26& 0.94& 1.03& 0.77& 1.11& 1.16\\
\ce{H2O}-\ce{F}&     4.25& 5.33& 5.87& -1.10& 1.75& 3.84& 0.50& 0.79& 0.86& 1.10& 1.21\\
\ce{H2O}-\ce{Li}&    1.58& 2.10& 2.36& 1.20& 1.92& 2.27& 1.32& 1.28& 1.35& 1.38& 1.32\\
\ce{H2O}-\ce{HNH2+}& -0.80& 0.12& 0.58& -0.73& 0.21& 0.71& -1.18& -1.18& -0.55& -1.05& -1.12\\
\ce{H2O}-\ce{NH3+}&  1.43& 2.22& 2.62& 0.06& 1.18& 2.04& 0.43& 0.54& 0.71& 0.66& 0.69\\
\ce{FH}-\ce{BH2}&    0.14& 0.51& 0.70& -0.04& 0.39& 0.69& 0.00& 0.04& 0.15& 0.08& 0.09\\
\ce{HF}-\ce{CO+}&   -5.07& -5.22& -5.30& 0.96& 2.08& 2.83& 0.35& 0.48& 1.44& 0.64& 0.64\\
\ce{FH}-\ce{NH2}&   -0.13 & 0.62 & 1.00 & -0.25 & 0.56 & 0.98 & -0.54 & -0.53 & -0.09 & -0.44 & -0.48\\
\ce{FH}-\ce{OH}&     0.27& 0.76& 1.00& 0.16& 0.69& 0.98& 0.01& 0.02& 0.28& 0.08& 0.05\\\hline
  \end{tabular}
  \caption{
The counterpoise corrected interaction energy errors (kcal/mol) of MP2, OOMP2, and their variants in the 13 data points in TA13.
  }
  \label{tab:ta13big}
  \end{sidewaystable}

In Table \ref{tab:ta13}, we present the statistical errors of MP2, OOMP2, and their variants on the TA13 set.
OOMP2 performs the best among the methods presented here. We note that $\omega$B97M-V has an RMSD of 2.75 kcal/mol, \cite{Mardirossian2017} a little worse than OOMP2.
OOMP2, SCS-OOMP2, and SOS-OOMP2 all improved the numerical performance compared to their parent methods.
Regularized OOMP2 methods perform very well and unscaled ones perform better than the scaled ones.
We also presented mean signed errors (MSEs) which are often used to infer a potential bias in statistical data.
The MSEs are all positive in Table \ref{tab:ta13} and would be smaller if we performed a TQ extrapolation along with counterpoise corrections.\cite{Halkier1999} 
\replacewith{The MSE of 0.01 kcal/mol in OOMP2 will likely become more negative in a larger basis benchmark.}{}
In summary, we found that regularization does not damage the performance of OOMP2 in describing non-bonded interactions in the TA13 set.
\revision{Overall, $\kappa$-OOMP2 performs the best in the TA13 set among those tested.}
\begin{table}
\includegraphics[scale=0.57]{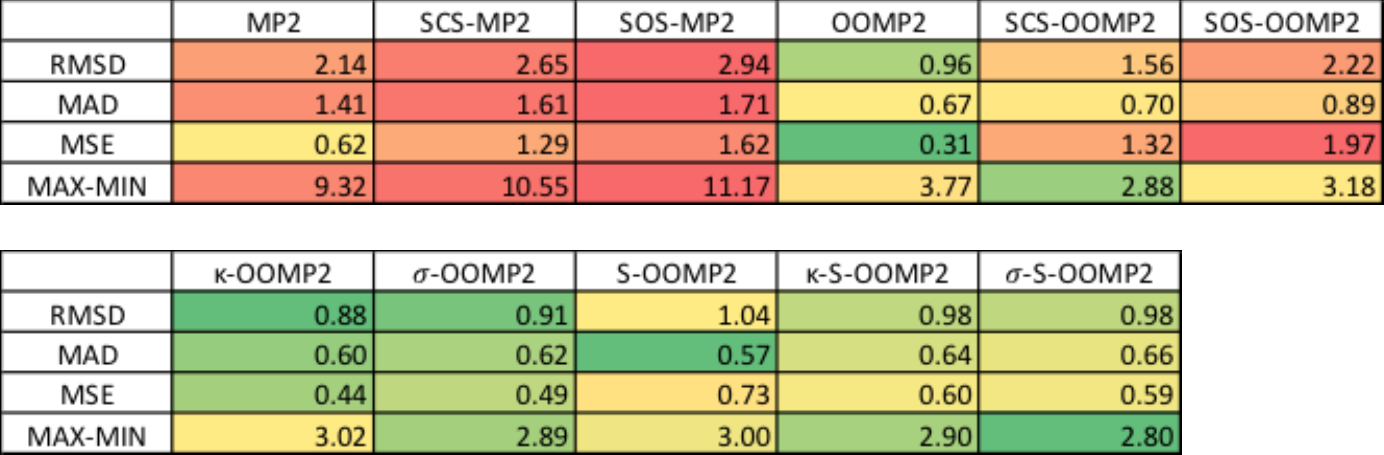}
\caption{\label{tab:ta13}
The TA13 set RMSD (kcal/mol)
of MP2, OOMP2, and their variants. MAD stands for Mean Absolute Deviation, MSE stands for Mean Signed Error, MAX-MIN stands for the difference between maximum and minimum. The colors represent the relative performance of each method; red means the worst and green means the best among the methods presented.
}
\end{table}
\subsection{Application to Organic Singlet Biradicaloids}
Organic biradicaloids are not very common to observe experimentally because they are quite unstable. 
\insertnew{Indeed, a molecule with a singlet biradical ground state is typically a contradiction. A singlet biradicaloid is the ground state due to the presence of some closed shell character.\cite{Jung2003}}
They may appear in numerous interesting chemical reactions as a transition state.\cite{Abe2013} In this section, we will study two experimentally observed organic singlet biradicaloids.\cite{Hu2015,Kobayashi2017}

One may attempt to use Yamaguchi's approximate spin-projected UMP2 (AP-UMP2) approach \cite{Yamaguchi1988, Yamaguchi1988a} to spin-project the broken-symmetry (BS) $M_{S}=0$ UMP2 state to obtain the spin-pure energy of the $S=0$ state. Assuming we have only singlet and triplet states that contribute to the $M_S=0$ state, one can easily work out the spin-pure singlet energy based on $\langle S^2 \rangle$:
\begin{equation}
E_{S=0}  = \frac{E_\text{BS} - (1- \alpha) E_{S=1}}{\alpha}
\label{eq:Eproj}
\end{equation}
where
\begin{equation}
\alpha = \frac{\langle S^2\rangle_{S=1} - \langle S^2 \rangle _\text{BS}}{\langle S^2\rangle_{S=1} - \langle S^2\rangle_{S=0}}
\end{equation}
The projection is exact only if there is only one spin-contaminant (i.e. the triplet state since we are interested in the singlet state). 
There are numerous ways to evaluate $E_{S=1}$ and $\langle S^2 \rangle_{S=1}$. We will choose the simplest way which is to assume
$E_{S=1} \approx E_{M_S=1}$ and replace $\langle S^2 \rangle_{S=1}$ with $\langle S^2 \rangle_{M_S=1}$. This requires $M_S=1$ calculations in addition to $M_S=0$ calculations.
For this reason UMP2 cannot be reliably applied to the singlet state ($S=0$) as the $M_S=0$ UHF state is often massively spin-contaminated.
The core orbitals are assumed to be more or less the same between ${M_S=0}$ and ${M_S=1}$. 

In passing, we note that more satisfying AP-UMP2 results may be obtained via the approach by Malrieu and co-workers which makes these assumptions exact in the case of biradicaloids.\cite{Coulaud2012,Coulaud2013,Ferre2015} This is achieved by allowing unrestriction only within the two electrons in two orbitals (2e,2o) active space with a possibility of using restricted open-shell formalism. Furthermore, a common set of core orbitals is used for the BS state and the $S=1$ state. 
\insertnew{Our group explored a similar approach called unrestricted in active pairs\cite{Lawler2010} which can be combined with the AP formula to produce a spin-pure energy.}

It is common to observe $\langle S^2\rangle_\text{BS} >> 1$ with a stable $M_S=0$ UHF solution of biradicaloids and thus it can be dangerous to apply the spin-projection. Moreover, the $M_S=1$ state tends to be also spin-contaminated in biradicaloids.  As a solution to this problem, one may try to use UOOMP2 to obtain minimally spin-contaminated $M_S=0, 1$ states. This  is not always possible due to the inherent numerical instability of UOOMP2 that commonly arises when applied to strongly correlated systems like biradicaloids. Indeed, for the biradicaloids studied here, we were not able to obtain the UOOMP2 energies due to this instability.

It is then natural to use regularized UOOMP2 to obtain the $M_S=0,1$ states of those systems. With the regularizers developed in this work, we do no longer run into the numerical instability. Therefore, the combination of regularized UOOMP2 and Yamaguchi's spin projection is quite attractive for simulating the electronic structure of biradicaloids. We note that AP-UOOMP2 is formally \insertnew{extensive as long as biradicaloism is not exceeded (i.e. spin-contamination is limited to a two-electron manifold).} 
%

In passing, we note that the first order correction to $\langle S^2\rangle$ for regularized OOMP2 can be obtained in the same way as the usual MP2 method.\cite{Lochan2007} The only difference for $\kappa$-OOMP2 is that we use regularized amplitudes instead of the unmodified ones. In the case of $\sigma$-OOMP2, we get a half contribution from the regularized ones and another half from the unregularized ones. This is obvious from the form of the modified TPDMs in each regularized OOMP2.

\subsubsection{Heptazethrene Dimer (HZD)}
\begin{figure}[h!]
\includegraphics[scale=0.5]{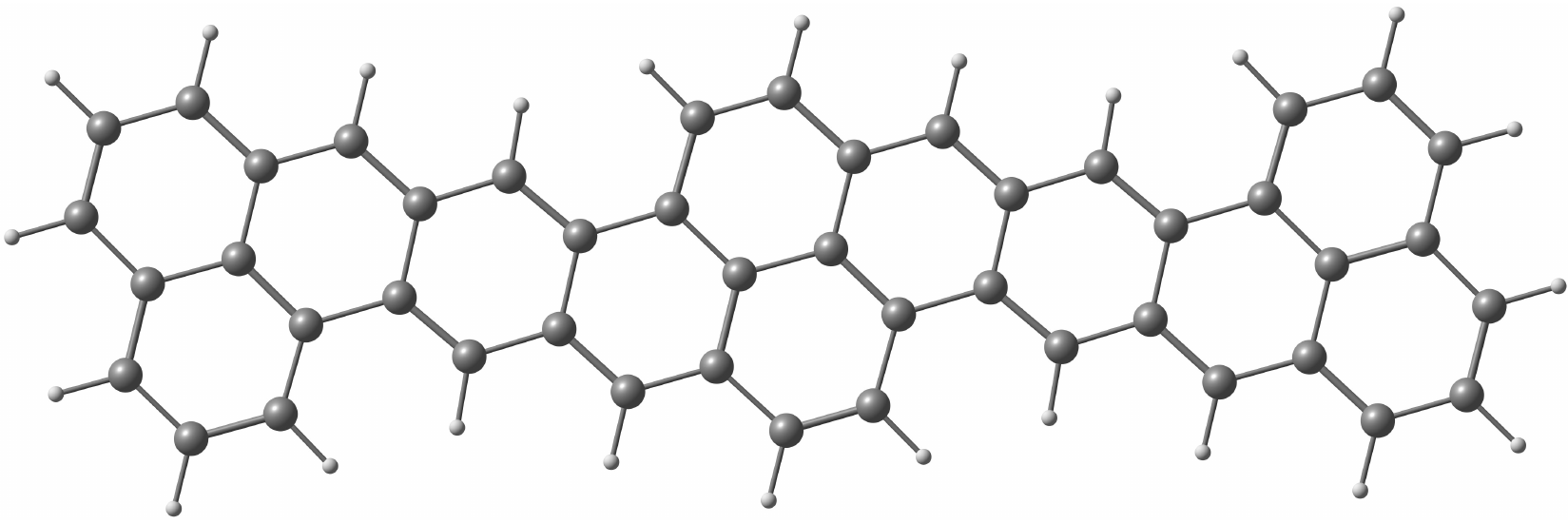}
\caption{\label{fig:hzd}
The molecular structure of heptazetherene dimer (HZD). White: H and Grey: C
}
\end{figure}

Oligozetherens have been experimentally synthesized and characterized as stable singlet biradicaloids. \cite{Hu2017} Similar to oligoacenes, they exhibit a polyradicaloid character in the background along with a strong biradicaloid character. There have been experimental interests in synthesizing tetraradicaloids using heptazethrenes. In particular, the experimental and theoretical work by Wu and co-workers has drawn our attention where they successfully synthesized heptazetherene dimer (HZD) as an attempt to synthesize a stable singlet tetraradicaloid. \cite{Hu2015} Using unrestricted CAM-B3LYP \cite{Yanai2004} density functional calculations, they characterized a strong biradicaloid character along with a very small tetraradicaloid character. They conclude that this compound should be better described as a biradicaloid and our work here also confirms this conclusion as we shall see.

The geometry was taken from ref. \citenum{Hu2015} and shown in Fig. \ref{fig:hzd}. We used the cc-pVDZ basis set \cite{Dunning1989} and the corresponding auxiliary basis set.\cite{Weigend2002} Furthermore, the frozen core approximation was employed to reduce the computational cost. 

\begin{table}
  \centering
  \begin{tabular}{r|c|r|r|r}\hline
\multicolumn{1}{c}{$M_S$}\vline 
& \multicolumn{1}{c}{$\kappa$-OOMP2} \vline
& \multicolumn{1}{c}{$\sigma$-OOMP2} \vline
& \multicolumn{1}{c}{$\kappa$-S-OOMP2} \vline
& \multicolumn{1}{c}{$\sigma$-S-OOMP2}
\\
\hline
0 & 0.00 (1.211) & 0.00 (1.145) &    0.00 (1.238) & 0.00 (1.173) \\
\hline
1 & 1.95 (2.117) & 1.62 (2.091) &    2.07 (2.129) & 1.74 (2.104) \\
\hline
2 & 46.07 (6.115) & 45.23 (6.091) & 45.99 (6.126) & 45.26 (6.104) \\
\hline
  \end{tabular}
  \caption{
  Spin-gaps (kcal/mol) of HZD from regularized UOOMP2 methods developed in this work.
  The numbers in parentheses are the corresponding $\langle S^2 \rangle$ value.
  }
  \label{tab:hzdenergy}
\end{table}

In Table \ref{tab:hzdenergy}, we present the spin gaps and $\langle S^2 \rangle$ of HZD using regularized UOOMP2 methods developed in this work. 
The gap between the $M_S=0$ and $M_S=1$ states is very small.
Furthermore, the $M_S=0$ state is heavily spin contaminated. 
This is a signature of biradicaloids. 
The $\langle S^2 \rangle$ values of the $M_S=1,2$ states are relatively close to the corresponding spin-pure states. There is also roughly  a gap of 45 kcal/mol between the triplet and the quintet state and this supports that HZD is not a tetraradicaloid and better described as a biradicaloid. Given those observations, this system is well suited for Yamaguchi's AP. Applying AP will yield a lower singlet state than the $M_S=0$ state and thus provide a larger singlet-triplet gap.

\begin{table}
  \centering
  \begin{tabular}{c|r|r|r|r}\hline
\multicolumn{1}{c}{}\vline 
& \multicolumn{1}{c}{$\kappa$-OOMP2} \vline
& \multicolumn{1}{c}{$\sigma$-OOMP2} \vline
& \multicolumn{1}{c}{$\kappa$-S-OOMP2} \vline
& \multicolumn{1}{c}{$\sigma$-S-OOMP2}
\\
\hline
$\alpha$ & 0.428 & 0.452 & 0.418 & 0.442\\
\hline
$\Delta E_{S-T}$ & 4.55 & 3.59 & 4.96 & 3.95\\
\hline
  \end{tabular}
  \caption{
  The spin-projected single-triplet gap $\Delta E_{S-T}$ (kcal/mol) of HZD from regularized AP-UOOMP2 methods.
$\alpha$ is the spin-projection coefficient used to obtain the projected energy in Eq. \eqref{eq:Eproj}.
  }
  \label{tab:hzdproj}
\end{table}

In Table \ref{tab:hzdproj}, the spin-projection coefficient $\alpha$ and the resulting spin-projected singlet-triplet gap are presented.
Different methods exhibit a different magnitude of $\alpha$ and $\Delta E_{S-T}$ and the range of $\Delta E_{S-T}$ is from 3.59 kcal/mol to 4.96 kcal/mol which is roughly a 1.4 kcal/mol variation. We also note that there is a roughly 1 kcal/mol difference between $\kappa$ and $\sigma$ regularizers in both unscaled and scaled variants. The scaled variants have a 0.5 kcal/mol larger $\Delta E_{S-T}$ compared to their corresponding unscaled variants. Regardless of which regularized OOMP2 we use, $\Delta E_{S-T}$ is small enough that this molecule is undoubtedly a biradicaloid. \insertnew{The extent of biradicaloid character can be inferred from the value of $\alpha$ in Table \ref{tab:hzdproj}. $\alpha=0.5$ is the perfect biradical limit and HZD shows $\alpha = 0.40-0.45$. This suggests that the stability of HZD may be attributed to some closed-shell configuration contribution.}

\subsubsection{Pentaarylbiimidazole (PABI) complex}
\begin{figure}[h!]
\includegraphics[scale=0.5]{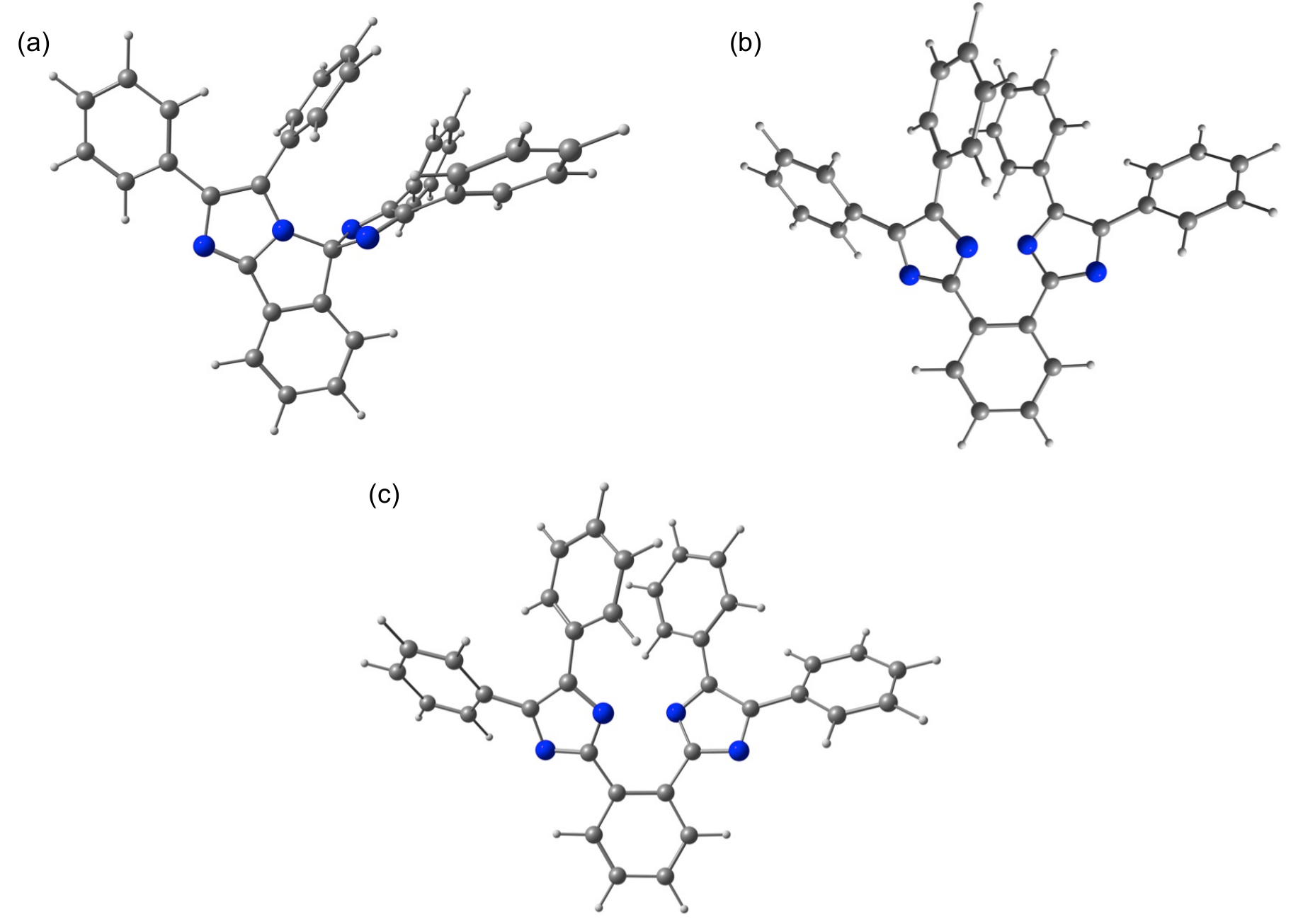}
\caption{\label{fig:pabi}
The molecular structures of PABI: (a) closed, (b) open 1 and (c) open 2. 
White: H, Grey: C, and Blue: N.
}
\end{figure}
Recently, Miyasaka, Abe, and co-workers studied a photochromic radical dimer, pentaarylbiimidazole (PABI) by a means of ultrafast spectroscopy. \cite{Kobayashi2017}
Without any external perturbations, PABI stays as its closed conformer shown in Fig. \ref{fig:pabi} (a). This stable conformation is closed-shell and does not exhibit any biradicaloid characters. Once a laser field is applied, the closed conformation undergoes a transition to its excited state and a subsequent relaxation back to the ground state surface. During this dynamics, the C-N bond in the middle in Fig. \ref{fig:pabi} (a) gets dissociated, which results in two possible conformers, open 1 and open 2, depicted in Fig. \ref{fig:pabi} (b) and (c), respectively. These two conformers were computationally shown \cite{Kobayashi2017} to exhibit quite strong biradicaloid characters, which drew our attention.

As understanding the dynamics in the system requires a reliable method for treating excited states and strongly correlated ground state, Miyasaka, Abe, and co-workers applied extended multi-state complete active space second-order perturbation theory (XMS-CASPT2). \cite{Shiozaki2011} Here we will compare AP-UOOMP2 ground state results against XMS-CASPT2 (4e, 6o).

All the geometries are obtained from ref. \citenum{Kobayashi2017} and we used def2-SVP \cite{Weigend2005} and the corresponding density-fitting basis \cite{Hattig2005} for the study of this molecule. The frozen core approximation was employed.

The closed conformation exhibits $\langle S^2 \rangle_0 = 3.31$ at the $M_S=0$ UHF level which is attributed to artificial symmetry breaking. To support this, we ran $\kappa$-S-OOMP2 and $\sigma$-S-OOMP2 with $\kappa$ and $\sigma$ ranging from 0.05 to 4.0. The scaling factors for each $\kappa$ and $\sigma$ are the optimal values which yielded Fig. \ref{fig:scsrmsd}. The $\kappa$ values greater than 0.2 and all values of $\sigma$ were enough to restore the restricted spin symmetry. This strongly suggests that the closed conformation is a closed-shell molecule with no strong correlation.

The open 1 and open 2 conformations are heavily spin-contaminated at the $M_S=0$ UHF level as they have $\langle S^2 \rangle_0 = 4.41$ and $\langle S^2 \rangle_0 = 4.47$, respectively. These two cases are particularly interesting because if the regularization strength is too weak then it fully restores the spin-symmetry and yields a closed-shell solution. $\kappa$-S-OOMP2 requires $\kappa$ less than 3.8 for open 1 and $\kappa$ less than 3.5 to observe spin-symmetry breaking. At $\kappa=1.5$ (recommended value), each conformer has $\langle S^2 \rangle=0.929$ and $\langle S^2 \rangle = 0.918$, respectively. This supports the conclusion that both open 1 and open 2 are biradicaloids as pointed in ref.\citenum{Kobayashi2017}. A similar result was found for $\sigma$-S-OOMP2 as well.

\begin{table}
  \centering
  \begin{tabular}{c|r|r|r|r}\hline
$\langle S^2\rangle$ & $\kappa$-OOMP2 & $\sigma$-OOMP2 & $\kappa$-S-OOMP2 & $\sigma$-S-OOMP2\\ \hline
open 1 ($M_S=0$) & 0.918 & 0.897 & 0.929 & 0.911\\ \hline
open 1 ($M_S=1$) & 2.026 & 2.023 & 2.030 & 2.026\\ \hline
open 2 ($M_S=0$) & 0.906 & 0.880 & 0.918 & 0.896\\ \hline
open 2 ($M_S=1$) & 2.027 & 2.023 & 2.031 & 2.027\\ \hline
  \end{tabular}
  \caption{
 The $\langle S^2\rangle$ values of regularized UOOMP2 methods.
  }
  \label{tab:pabis2}
\end{table}

In Table \ref{tab:pabis2}, we present the $\langle S^2\rangle$ values of the $M_S=0, 1$ states of open 1 and open 2. Different flavors of regularized OOMP2 do not deviate significantly from each other. The $M_S=1$ states are all almost spin-pure whereas the $M_S=0$ states are spin-contaminated. As the $\langle S^2\rangle$ values of the $M_S=0$ states are close to 1.0, those states exhibit significant biradicaloid character. Therefore, these two systems are perfect candidates for the Yamaguchi's AP scheme.

\begin{table}
  \centering
  \begin{tabular}{c|r|r|r|r|r}\hline
\multicolumn{1}{c}{}\vline  & XMS-CASPT2$^1$ & $\kappa$-OOMP2 & $\sigma$-OOMP2 & $\kappa$-S-OOMP2 & $\sigma$-S-OOMP2\\
 \hline
closed & 0 & 0 & 0 & 0 & 0\\
\hline
open 1 & 32.03 & 37.13 & 36.52 & 36.21 & 35.97\\
\hline
open 2 & 33.57 & 38.79 & 38.11 & 37.78 & 37.50\\
\hline
  \end{tabular}
  \caption{
  The relative energies (kcal/mol) of the three conformers of PABI from XMS-CASPT2 and regularized AP-UOOMP2 methods.
$^1$ The XMS-CASPT2 numbers were taken from ref. \citenum{Kobayashi2017} and the active space used was (4e, 6o). 
  }
  \label{tab:pabiproj}
\end{table}

Applying the AP scheme to spin-purify the spin-contaminated $M_S=0$ energies of open 1 and open 2 leads to the various OOMP2 relative energies shown in Table \ref{tab:pabiproj}. The results in Table \ref{tab:pabiproj} show almost no quantitative differences between different regularized OOMP2 methods. It is interesting that the system is quite insensitive to what flavor of OOMP2 we use. Compared to XMS-CASPT2, the relative energies of regularized OOMP2 for open 1 and open 2 are roughly 5 kcal/mol higher. The relative energies between open 1 and open 2 are reproduced by every regularized OOMP2 presented: open 2 is about 1.5 kcal/mol higher than open 1. While the regularized OOMP2 methods agree with XMS-CASPT2 on a very small relative energy scale between open 1 and open 2, they differ significantly from XMS-CASPT2 for the relative energy between closed and open conformations.  It is unclear whether this small active space XMS-CASPT2 is a reliable method for this problem \insertnew{just as it is unclear that regularized OOMP2 is quantitatively accurate}. This is an interesting system to further study using a recently developed couple-cluster method in our group that can handle a much larger active space. \cite{Lee2017} 

In summary, we applied the regularized OOMP2 methods developed in this work to obtain relative energies of three conformations of PABI. Two of the three conformations were found to be strong biradicaloids which agree well with what was found with the XMS-CASPT2 study before. We also found that the different regularized OOMP2 methods do not differ significantly from each other.

\section{Conclusions}

Orbital-optimized second order M{\o}ller-Plesset perturbation theory (OOMP2) is an inexpensive approach to obtaining approximate Br{\"u}ckner orbitals, and thereby cutting the umbilical cord between MP2 and mean-field Hartree-Fock (HF) orbitals. 
This has demonstrated benefits for radicals and systems where HF exhibits artificial symmetry-breaking. \cite{Lochan2007}
However the limited MP2 correlation treatment can introduce artifacts of its own, because the MP2 correlation energy diverges as the HOMO-LUMO gap approaches zero. One striking example is that restricted and unrestricted orbital solutions are each local minima for molecules with stretched bond-lengths -- in other words there is no Coulson-Fischer point \cite{Coulson1949,Sharada2015} where the restricted orbital solution becomes unstable to spin polarization! It has been previously recognized that some type of regularization is necessary to avoid such divergences. Simple level shifts have been explored,\cite{Stuck2013} but are inadequate in general because the size of the level shift needed to ensure a Coulson-Fischer point in general is so large that the MP2 correlation effects are grossly attenuated.\cite{Razban2017}

In this work we have therefore developed and assessed two new regularization approaches, called $\kappa$-OOMP2 and $\sigma$-OOMP2, which both have the feature that the strength of regularization is largest as the HOMO-LUMO gap approaches zero, and becomes zero as the gap becomes large. This way the total correlation energy is not greatly attenuated even with quite strong regularization. The regularization strength in each case is controlled by a single parameter (i.e. $\kappa$ and $\sigma$) which has units of inverse energy so that small values correspond to strong regularization. Despite the greater complexity of these regularizers relative to a simple level shift, they can be quite efficiently implemented in conjunction with orbital optimization, at a cost that is not significantly increased relative to unmodified OOMP2. These models can be used with just the single parameter, or, alternatively, an additional parameter corresponding to scaling the total correlation energy (i.e. S-OOMP2) can be included as well.

The main conclusions from the numerical tests and assessment of the regularizers are as follows: 

\begin{enumerate}

	\item{\textit{Regularization}. We assessed the performance of the new regularizers on single, double and triple bond-breaking problems, to determine the weakest regularizers that can properly restore the Coulson-Fischer (CF) points across these systems. The conclusion is very encouraging: a regularization parameter of $\kappa \le 1.5$ $E_{\textrm{h}}^{-1}$ is capable of correctly restoring the CF points on all of these systems. For $\kappa = 1.5$ $E_{\textrm{h}}^{-1} $ regularization applied the ethane, ethene, ethyne series, the CF distance, $r_{\textrm{CF}}$ is much shorter for the two latter systems as is appropriate for the physics of the method. The $\sigma$ regularization is clearly less satisfactory in this regard, as $r_{\textrm{CF}}$(C$_2$H$_2) > r_{\textrm{CF}}($C$_2$H$_4$) for the smallest $\sigma=1.0$ $E_{\textrm{h}}^{-1}$ value considered, and the lowest eigenvalue of the stability matrix does not always show monotonic behavior as a function of bond-stretching displacements. }
	
	\item{\textit{Scaling}. We examined the performance of OOMP2 with and without regularizers, as well as with and without scaling (S) of the total correlation energy on two datasets representing thermochemical energy differences (W4-11) and radical stabilization energies (RSE43) \insertnew{and one dataset representing radical-closed-shell non-bonded interaction energies (TA13)}.  The TAE140 subset of the W4-11 set was used to train scaling factors. The results show that unregularized OOMP2 over-emphasizes correlation effects, as the optimal scaling factor is only 0.9. By contrast, choosing a physically appropriate $\kappa$ value of 1.45 $E_{\textrm{h}}^{-1}$ is appropriate for use without scaling, by reducing the tendency of orbital optimization to over-correlate through smaller energy gaps.  A slight improvement in numerical results can be obtained with a scaled $\kappa$-S-OOMP2 method, using $\kappa = 1.5$ $E_{\textrm{h}}^{-1}$ and $c=0.955$. Broadly similar conclusions hold for $\sigma$ regularization.
\revision{The regularized OOMP2 methods perform slightly better than OOMP2 for the TA13 set, and slightly worse for the RSE43 set.}
	}
	
	\item{\textit{Chemical application to singlet biradicaloids}. 
We applied these regularized OOMP2 methods to two experimentally relevant organic biradicaloids, the heptazethrene dimer (HZD) and the pentaarylbiimidazole complex (PABI). We emphasize that unmodifed OOMP2 diverges for these systems and the regularization is necessary to obtain energies in a numerically stable way. We combined the regularized OOMP2 methods with Yamaguchi's approximate projection scheme to spin-purify $M_S=0$ energies of the biradicaloids. We found that all four regularized OOMP2 methods developed in this work perform equally well.
	}
	
	\item{\textit{Recommendation}. Given the documented failures of OOMP2 for bond-breaking without regularization, and its related tendency to over-correlate, it cannot be recommended for general chemical applications despite its formal advantages. Fortunately, the $\kappa = 1.45$ $E_{\textrm{h}}^{-1}$ regularization introduced here appears to resolve all of these issues in a way that is as satisfactory as could be hoped for, given that MP2 itself is inherently incapable of solving strong correlation problems (i.e. spin-polarization should occur in such cases). We recommend $\kappa$-OOMP2 as a more robust replacement for OOMP2. We believe that it may also be valuable as a way of realizing well-behaved orbital optimized double hybrid density functionals\cite{Peverati2013,Najibi2018} in the future.}

\end{enumerate}
\revision{Beyond stabilizing the OOMP2 method, the new regularizers introduced in this work may also have other interesting and potentially useful applications in electronic structure theory. For example, they can be applied to M{\o}ller-Plesset theory without orbital optimization. At the MP2 level this will alter the relative energies of RMP2 and UMP2 in a way that raises the RMP2 energy when energy gaps are small, possibly avoiding artifacts that occur in that regime. It may also be interesting to explore the effect on higher order correlation energies, such as MP3 or MP4, or the triples correction to methods such as coupled cluster theory with singles and doubles, CCSD(T). Orbital optimization can also be performed with coupled cluster doubles (i.e. OO-CCD),\cite{Krylov1998} and for cases where electron correlation effects are strong, regularization may be also be useful to ensure the correct presence of Coulson-Fischer points. Likewise regularizers may be helpful to avoid non-variational failures of coupled cluster theory without orbital regularization. Of course, it is an open question whether the forms we have presented here are appropriate for these non-MP2 applications or not.}

\newpage

\section{Table of Contents Graphic}
\begin{figure}[h!]
\includegraphics[scale=0.65]{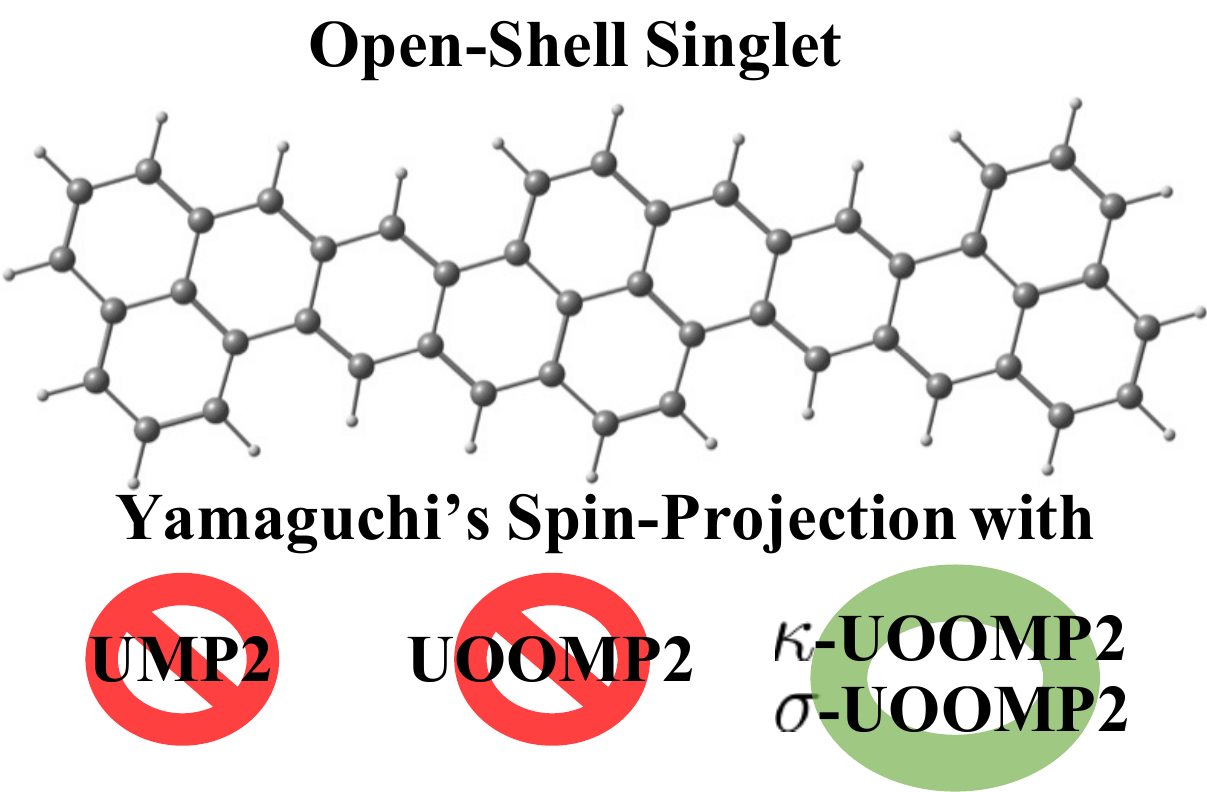}
\end{figure}

\section{Acknowledgement}
J. L. thanks 
Evgeny Epifanovsky for useful discussions on the efficient implementation, Narbe Mardirossian, and Yuezhi Mao for useful suggestions and discussions on the regularization parameter training, David St{\"u}ck, Roberto Peverati, and Eloy Ramos for some previous attempts related to this work and Soojin Lee for consistent encouragement and support.
We are grateful to Luke Bertels for validating the MP2 data reported for the W4-11 and TA13 data sets.
This work was supported by a subcontract from MURI Grant W911NF-14-1-0359.

\bibliography{uoomp2_ms_rev1}
\bibliographystyle{achemso}
\end{document}